\documentclass[prc,twocolumn,showpacs,superscriptaddress,preprintnumbers,amsmath,amssymb,nofootinbib]{revtex4}

\usepackage[dvipdfmx]{graphicx}
\usepackage{dcolumn}
\usepackage{bm}
\usepackage{amsmath}

\def\GeV{\:{\rm GeV}}
\def\sNN{\sqrt{s_{NN}}}
\def\pT{p_T}

\usepackage[normalem]{ulem}  
\usepackage[dvips]{color} 

\renewcommand\sout{\bgroup \color{red} \ULdepth=-.5ex \ULset}

\begin{document}

\title{
Effects of hadronic rescattering on multistrange hadrons\\
in high-energy nuclear collisions
}

\author{Shiori Takeuchi}
\email{takeuchi@sophia.ac.jp}
\affiliation{%
Department of Physics, Sophia University, Tokyo 102-8554, Japan
}
\author{Koichi Murase}
\email{murase@nt.phys.s.u-tokyo.ac.jp}
\affiliation{Department of Physics, The University of Tokyo, Tokyo 113-0033, Japan}
\affiliation{%
Theoretical Research Division, Nishina Center, RIKEN, Wako 351-0198, Japan
}
\affiliation{%
Department of Physics, Sophia University, Tokyo 102-8554, Japan
}
\author{Tetsufumi Hirano}
\email{hirano@sophia.ac.jp}
\affiliation{%
Department of Physics, Sophia University, Tokyo 102-8554, Japan
}
\author{Pasi Huovinen}
\email{huovinen@th.physik.uni-frankfurt.de}
\affiliation{%
Institut f{\"u}r Theoretische Physik, Goethe-Universit{\"a}t, 60438 Frankfurt am Main, Germany
}
\affiliation{%
Frankfurt Institute for Advanced Studies, 60438 Frankfurt am Main, Germany
}
\author{Yasushi Nara}
\email{nara@aiu.ac.jp}
\affiliation{%
Department of International Liberal Arts, Akita International University, Yuwa, Akita-city 010-1292, Japan
}

\date{\today}

\begin{abstract}
We study the effects of hadronic rescattering on hadron distributions
in high-energy nuclear collisions
by using an integrated dynamical approach.
This approach is based on a hybrid
model combining (3+1)-dimensional
 ideal hydrodynamics
for the quark gluon plasma (QGP),
 and a transport model for the hadron resonance gas.
Since the hadron distributions are the result of the entire expansion
history of the system, understanding the QGP properties requires
investigating how rescattering during the hadronic stage affects the
final distributions of hadrons.
We include multistrange hadrons in our study, and quantify the
effects of hadronic rescattering on their mean transverse momenta and
elliptic flow.  We find that multistrange hadrons scatter less during
the hadronic stage than non-strange particles, and thus their
distributions reflect the properties of the system in an earlier stage
than the distributions of non-strange particles.
\end{abstract}


\maketitle

\section{Introduction}
Quark gluon plasma (QGP),
strongly interacting matter composed of quarks and gluons, 
has been created in high-energy nuclear collisions 
at the Relativistic Heavy Ion Collider (RHIC) at Brookhaven
National Laboratory (BNL)
and the Large Hadron Collider (LHC) at CERN~\cite{Yagi:2005yb}.
One of the main discoveries at the RHIC
was that the QGP behaves like a nearly perfect 
fluid~\cite{Heinz:2001xi,Gyulassy:2004vg,sQGP1,sQGP2,sQGP3,Hirano:2005wx},
and since then the transport properties of the QGP, especially its
shear and bulk viscosity coefficients, have been under active
investigation.

The QGP created in high-energy nuclear collisions
expands, cools down, 
and turns into a hadron gas.
Since hadrons scatter with each other via the strong interaction,
information about the QGP stage
is, in general, contaminated by rescatterings in the hadronic stage.
To probe the QGP more directly,
thermal photons and dileptons
were proposed as penetrating probes~\cite{Feinberg:1976ua,Shuryak:1978ij}.
Once emitted, photons and dileptons
propagate to detectors without rescattering
since they interact only electromagnetically
and their mean free path is therefore longer than the typical size of the
system. However, photons and dileptons 
are not perfect probes of the QGP either.
They are emitted during all the stages of high-energy nuclear collisions 
such as the primary hard scatterings, pre-equilibrium stage, 
 and late hadronic stage~\cite{Bratkovskaya:2014mva}, and it is not possible
to distinguish 
photons and dileptons emitted from the QGP from those
originating from the other processes.

In this paper, we follow up the previous
studies~\cite{Shor:1985prl,vanHecke:1998prl,Bass:2000ib,Cheng:2003prc,Hirano:2008prc,He:2011zx,Zhu:2015dfa}
about multistrange hadrons, and claim that their distributions allow
to probe the QGP immediately after hadronization.
Multistrange hadrons, in particular $\phi$ mesons and $\Omega$ baryons,
have small scattering cross-sections with pions since they do not form any
resonances unlike other hadrons.
This is a unique property of the multistrange hadrons:
Their distributions reflect the properties of the system
mainly at a specific stage deep inside the fireball,
unlike photons and dileptons
which come from the all stages.

By using a hybrid model in which hydrodynamic description of the QGP
fluid is followed by a hadronic cascade model, 
violation of the mass ordering of the differential elliptic flow parameter
$v_2(p_T)$ was predicted~\cite{Hirano:2008prc}. 
This phenomenon was observed recently
by the STAR Collaboration \cite{Nasim:2013npa}, 
which indicates that $\phi$ mesons are less affected
by hadronic rescatterings and, consequently, that
they are good probes deep inside the matter.
In this paper, we investigate the hadronic rescattering effects on
observables more systematically by employing an integrated dynamical
approach which we have previously used to analyze various observables
at the RHIC and the LHC energies~\cite{Hirano:2013ppnp}.
In the previous study of the violation of mass ordering~\cite{Hirano:2008prc}, 
a first order phase transition model for the equation of state (EoS)
 was employed  
and only hadrons up to the mass
of  $\Delta(1232)$ were taken into account
in the hadron phase.
In the present paper, we use a more realistic equation of
state~\cite{Huovinen:2009yb} connecting a parametrized lattice QCD EoS
at large temperatures to a hadron resonance gas EoS at low
temperatures. The hadron resonance gas in the EoS contains all the
resonances in the hadron cascade model JAM (Jet $AA$ Microscopic transport model)~\cite{Nara:1999dz} used as
the last stage of the integrated dynamical approach.
Hence, in addition to $\phi$, we are able to investigate
hadronic rescattering effects on heavier multistrange hadrons
such as $\Xi$ and $\Omega$.

In the following we introduce the integrated dynamical model in
Sec.~\ref{sec:model}. We compare transverse momentum ($\pT$) spectra
and $\pT$-differential elliptic flow parameters $v_{2}(\pT)$ with the
STAR data in Sec.~\ref{sec:results}.  After we confirm we reasonably
reproduce the data by using this model, we discuss the hadronic
rescattering effect in Sec.~\ref{sec:RescatteringEffects}. In this
section, we first revisit the violation of mass ordering of
$v_{2}(\pT)$.  We next focus on the mean transverse momentum 
$\langle \pT \rangle$ and $p_T$-averaged $v_{2}$ calculated with or without hadronic
rescatterings and investigate how much these final observables reflect
the information when the hydrodynamic stage finishes. We summarize our
results in Sec.~\ref{sec:summary}.

\section{The Model}
\label{sec:model}

We describe the space-time evolution of high-energy nuclear collisions by
an integrated dynamical approach on an event-by-event basis.
We divide the whole reaction
into three 
separate stages: Initial stage, hydrodynamic stage, and transport stage.
The initial entropy density distribution after the collision of energetic
heavy nuclei
is parametrized by using a Monte-Carlo version of the Glauber model.
The subsequent expansion of the matter is described by
relativistic ideal hydrodynamics. 
When the system becomes sufficiently dilute due to strong expansion,
we switch the description of the system from hydrodynamics to kinetic theory.

In the following we briefly overview each part of the model and the
corresponding interfaces. For further details, see
Ref.~\cite{Hirano:2013ppnp}.

\subsection{Hydrodynamics and equation of state}

Hydrodynamics is a macroscopic description and an effective theory of
the system's long wavelength/time behavior.
The hydrodynamical equations of motion are obtained from the
conservation of energy and momentum:
\begin{equation}
\label{eq:hydro}
\partial_{\mu} T^{\mu \nu} = 0,
\end{equation}
where $T^{\mu \nu}$ is energy-momentum tensor.
In the ideal fluid approximation,
the energy momentum tensor can be decomposed as
\begin{equation}
\label{eq:ideal}
T^{\mu \nu}  = e u^{\mu} u^{\nu} - P (g^{\mu \nu} - u^{\mu} u^{\nu}),
\end{equation}
where $e$, $P$, $u^{\mu}$, and $g^{\mu \nu} = \mathrm{diag}(+1, -1, -1, -1)$
are energy density, pressure, flow velocity, and the Minkowski metric,
respectively.
To close the system of partial differential equations we need to know
the equation of state (EoS) of the fluid to obtain pressure $P$ as a
function of energy density $e$. Once the EoS is known, the space-time
evolution of thermodynamic quantities and flow velocity is determined
for given initial conditions, i.e., energy density distribution and
flow field at initial time $\tau_0$.

We ignore the baryon number and its continuity equation since at the
collider energies, net baryon density at midrapidity is tiny.
When we solve Eq.~(\ref{eq:hydro}), temperature
is not needed since it is merely a parameter to connect energy density
and pressure.
Nevertheless, we also follow the space-time evolution of temperature
to decide where to switch from fluid to cascade.
We employ the \textit{s95p}-v1.1 parametrization for the EoS. The
\textit{s95p} parametrization~\cite{Huovinen:2009yb} connects the
lattice QCD based EoS~\cite{Bazavov:2014prl} in the high temperature
region to a hadron resonance gas EoS in the low temperature region.
The hadronic part of \textit{s95p}-v1.1 contains the same hadronic
species as the JAM hadronic cascade model~\cite{Nara:1999dz} described
in the next subsection.

We numerically solve Eq.~(\ref{eq:hydro}) in the Milne coordinates 
$(\tau, \eta_{s}, x, y)$
which is appropriate for the description of the evolution 
at relativistic energies \cite{Hirano:2001eu}.
Here $\tau = \sqrt{t^2 -z^2}$ is longitudinal proper time,
$\eta_{s} = \frac{1}{2}\ln[(t+z)/(t-z)]$ is space-time rapidity,
and $x$ and $y$ are transverse
coordinates perpendicular to the collision axis.
In the numerical calculations, we employ the Piecewise Parabolic Method 
\cite{Colella:1982ee}
which is known to be able to describe shock waves.
For details about this numerical algorithm, see also Ref.~\cite{Hirano:2012yy}.

\subsection{Particlization and hadronic cascade}

We employ the hadronic cascade model JAM \cite{Nara:1999dz} to describe the space-time
evolution of hadron gas after  ``particlization", i.e., after switching from fluid to particles.
At the late stage of collisions, the system is too  dilute  to maintain equilibrium. 
We assume 
this happens around temperature, 
which we call the switching temperature $T_{\mathrm{sw}}$,  
and  switch description from hydrodynamics to kinetic theory 
on $T = T_{\mathrm{sw}}$ isosurface
\cite{Bass:1999tu,Bass:2000ib,Teaney:2000cw,Teaney:2001av,Hirano:2005xf,Nonaka:2006yn,Petersen:2008dd}.
Within the current model,
$T_{\mathrm{sw}}$ is rather adjustable parameter
which controls the final particle ratios.
We choose  $T_{\mathrm{sw}} = 155$ MeV 
to reproduce the observed 
$p_{T}$ spectra of pions, kaons, and protons + anti-protons 
in the low $p_{T}$ region \cite{Hirano:2013ppnp}
at the full RHIC energy.
At $T(x) = T_{\mathrm{sw}}$,
we calculate the single particle phase space distributions $f_{i}(p,x)$ 
for all hadrons included in the EoS
by employing the Cooper-Frye prescription~\cite{Cooper:1974mv}
\begin{equation}
\label{eq:CF}
f_{i}(p, x) d^{3}\bm{x} d^{3}\bm{p}= \frac{g_{i}}{(2 \pi)^3 E} \frac{p \cdot \Delta\sigma d^{3}\bm{p}}{\exp(p\cdot u/T_{\mathrm{sw}}) \pm 1}.
\end{equation}
Here $g_{i}$ is a degeneracy of the hadronic species $i$, $p^\mu = (E, \bm{p})$ 
is four-momentum of a particle, and
$\Delta \sigma_\mu$ is a normal vector of the $T(x) = T_{\mathrm{sw}}$
hypersurface.
It is well known that at some momenta the Cooper-Frye formula (\ref{eq:CF})
gives negative contribution for 
space-like hypersurface element (or even time-like hypersurface element with
negative time component).
These negative numbers correspond to in-coming particles,
which one cannot treat in the hadronic cascade model.
So we just neglect these contributions and consider only out-going particles.
We sample particles by going through all the hypersurface elements and generate
ensembles of hadrons on an event-by-event basis.
Note that we do not oversample particles to gain statistics for each event. This means
each hydrodynamic event corresponds to one ensemble of particles.
Among all the calculations in the integrated dynamical approach,
sampling particles from particlization hypersurface
is numerically most expensive, which is crucial in event-by-event simulations.
In Ref.~\cite{Hirano:2013ppnp},
 we discussed in detail  how to sample particles and
 integrate the Cooper-Frye formula 
(\ref{eq:CF}) over momentum at less numerical cost. 

In hadronic cascade models,
experimental hadronic cross section data 
are implemented when available.
However, there are many hadronic scattering processes where data do not exist.
In such a case, we use the additive quark model \cite{rqmd2,rqmd4,urqmd2,urqmd3}
to describe the scattering cross section
\begin{equation}
\sigma_{\mathrm{tot}}^{12} = \sigma_{\mathrm{tot}}^{NN} \frac{n_{1}}{3} \frac{n_{2}}{3}
\left(1-0.4 \frac{n_{s1}}{n_{1}} \right)\left(1-0.4 \frac{n_{s2}}{n_{2}} \right).
\end{equation}
Here $\sigma_{\mathrm{tot}}^{NN}$ is the total nucleon-nucleon cross section,
$n_{i}$ is the number of constituent quarks in a hadron,
and $n_{si}$ is the number of strange quarks in a hadron.
Thus hadrons containing strange quark(s) have relatively small total
cross section in this model.
Moreover, $\phi$ mesons and $\Omega$ baryons do not form resonances
when scattering with pions and therefore have much smaller cross sections than
non-strange hadrons.

In the simulations, dynamics of the system is described by a sum of incoherent
two-body collisions.
Two particles collide with each other if their minimum distance $b$
in the center of mass frame is smaller than
the distance given by the total cross section 
\begin{equation}
b < \sqrt{\frac{\sigma_{\mathrm{tot}}}{\pi}}.
\end{equation}
This is a classical, geometrical interpretation of the 
two-body collision cross section.

Note that in JAM the properties of hadrons are those of free
particles. The medium modifications of hadron properties such as a
broadening and mass shift{~\cite{Ko:1997kb}} have been extensively
studied by various further developments of transport models
such as the relativistic Vlasov-Uehling-Uhlenbeck (RVUU)
model~\cite{Fang:1994cm},
the hadron-string-dynamics (HSD~\cite{Cassing:1999es}) and
parton-hadron-string-dynamics (PHSD~\cite{Cassing:2008nn}) transport
models, the Giessen Boltzmann-Uehling-Uhlenbeck (GiBUU)~\cite{Buss:2011mx},
and the Isospin Quantum Molecular Dynamics (IQMD)~\cite{Hartnack:2011cn}.
Nevertheless, since the main difference in the cross sections of
multistrange particles and the other hadrons is due to the
(non-)existence of resonance channels, we do not expect the in-medium
modifications to significantly affect this difference, and the use of
free particle properties to provide a conservative baseline is
justified.

As mentioned, the interactions in JAM are described as two particle
scatterings, and multi-particle scatterings are not included. It has
been argued that multi-pion fusion processes in particular would have
a significant effect on the yields of proton-antiproton
pairs~\cite{Rapp:2000gy}, but subsequent calculations have shown that
regeneration is a $\sim 10$\% level effect at the
LHC~\cite{Steinheimer:2012rd,Pan:2014caa}. Nevertheless, the
multi-meson fusion effects for $\Omega$ baryons are much smaller than for
protons or $\Lambda$~\cite{Huovinen:2003sa}, and thus their inclusion
would make the later-discussed differences between rescatterings of
non-strange and multistrange particles even larger.

As a default setting 
 strong interactions and decays are simulated in JAM.
However, we switch off the decay channels for $\phi$ mesons 
in the present study to investigate hadronic rescattering effects on $\phi$
mesons in an efficient way.
Otherwise, we would need to perform mass reconstruction of $\phi$ mesons
from two kaons after numerical simulations.
Since the lifetime of $\phi$ mesons ($\sim 47$ fm/$c$) \cite{PDG:2014CPC}
is somewhat larger than the typical life time of the system ($\sim 10$ fm/$c$),
the daughter kaons from $\phi$ meson decays (or lack of them)
are not expected to affect the bulk evolution.
As we will discuss in Sec.~\ref{sec:results}, 
the yield of $\Lambda$ measured by
STAR contains the feed down from electromagnetic $\Sigma^{0}$ decays.
This process does not happen in the default setting in JAM
and, therefore, is simulated separately when we analyze $\Lambda$ spectra.
We also have an option in JAM to deactivate all the hadronic rescatterings
and/or resonance decays to investigate how these affect
final observables.

\subsection{Initial conditions}

Once initial conditions for the hydrodynamic evolution are fixed, the
subsequent evolution is determined, and we are able to obtain final
particle distributions which can be compared with experimental data.
The pre-thermalization stage between the first contact of colliding nuclei
and the initial time of hydrodynamic evolution
may be described by non-equilibrium field theory, although
how exactly it could be done is
one of the open issues in the physics of relativistic heavy ion collisions.
Thus, we do not even try to describe it and
simply rely on the Monte-Carlo (MC) Glauber~\cite{Miller:2007ri} 
and the modified Brodsky-Gunion-Kuhn (BGK) models~\cite{Hirano:2005xf}
to give the density distribution at $\tau_0$ in the transverse plane and
in the longitudinal direction, respectively.
The MC Glauber model enables us to calculate the number density of
participants and of binary collisions on an event-by-event basis.
Our assumption is that the initial entropy density
is proportional to a linear combination of these two densities.
Even at very high collision energies matter profile in the longitudinal
  direction depends on space-time rapidity, and boost invariance does not
  hold. Furthermore, in non-central collisions the longitudinal density
  gradient at mid space-time rapidity may be non-zero due to the different
  thicknesses of the colliding nuclei at each transverse position.
This can be deduced from (pseudo-)rapidity distribution of hadrons in 
p(d)-A collisions.
In the  modified BGK model smooth
initial longitudinal profile is parametrized by taking account
of the difference in the local thickness of colliding nuclei.
It is noted that
we do not employ the Monte-Carlo 
Kharzeev-Levin-Nardi
model~\cite{Drescher:2006ca,Drescher:2007ax}
which was used to calculate initial conditions in our previous
hydrodynamic studies~\cite{Hirano:2013ppnp}
since, in the present study, we focus on
the hadronic rescattering effects on transverse dynamics,
which are not supposed to be sensitive to the choice of initial conditions.

Assuming local thermalization,
energy density distribution $e(\tau=\tau_{0}, \eta_{s}, x, y)$ and pressure 
$P(\tau=\tau_{0}, \eta_{s}, x, y)$
are obtained from the
entropy density distribution through the \textit{s95p}-v1.1 EoS.
Initial  flow velocity at $\tau_{0}$ is supposed to be 
Bjorken's scaling flow \cite{Bjorken:1982qr}, namely,
$u^{\tau} = 1$ and $u^{x} = u^{y} = u^{\eta_{s}} =0$.
Throughout this study, we fix $\tau_0 = 0.6$ fm/$c$. 
We choose initial parameters to reproduce the
$p_{T}$ distributions of pions, kaons, and protons + anti-protons
in Au+Au collisions at $\sqrt{s_{NN}} = 200$ GeV
measured by the PHENIX Collaboration \cite{Adler:2003cb}.
The centrality dependence of multiplicity is controlled by
a fraction of soft (participants) and hard (binary collisions)
component in the MC Glauber model, whereas the multiplicity in central
collisions is fixed by the overall normalization parameter of the
initial entropy distribution. On the other hand, as mentioned
particle ratios are controlled by the switching temperature. 
The parameters controlling the shape of the rapidity distribution of
particles are kept the same as in our previous
study~\cite{Hirano:2013ppnp}. Since here we concentrate on observables
at midrapidity, we do not discuss them further.

In hydrodynamic simulations, one needs to convert the position
distribution (a sum of delta functions in the transverse plane) of
collision points in MC Glauber model to a hydrodynamic density
distribution. In our model we obtain the density at a point $x$ by
counting the number of collision points or participants within radius
$r_0 = \sqrt{\sigma_{\mathrm{in}}/\pi}$ around the point, and divide
by the inelastic cross section in $p$+$p$ collisions at
$\sqrt{s_{NN}}=200$ GeV, $\sigma_{\mathrm{in}}$.  For further details,
see Ref.~\cite{Hirano:2013ppnp}.

\section{Spectra and flow}
\label{sec:results}

In this section, we compare invariant $\pT$ spectra
and differential elliptic flow parameter $v_{2}(p_{T})$ 
of identified hadrons with experimental data from the STAR Collaboration.
In the present study, we analyzed 0.6 million ``minimum bias'' events
 (defined as $N_{\mathrm{part}} \ge 2$ in the MC Glauber calculations for
hydrodynamic initial conditions).
Centrality is defined using an event distribution of 
the charged hadron multiplicity
in $|\eta |< 0.5$ as done by the STAR Collaboration~\cite{Adler:2002xw}. 
Note that error bars in the theoretical plots denote the statistical errors.

\subsection{Transverse momentum spectra} 
\label{subsec:spectra}

\begin{figure*}[tbp]
\begin{center}
\includegraphics[clip,angle=-90,width=0.46\textwidth]{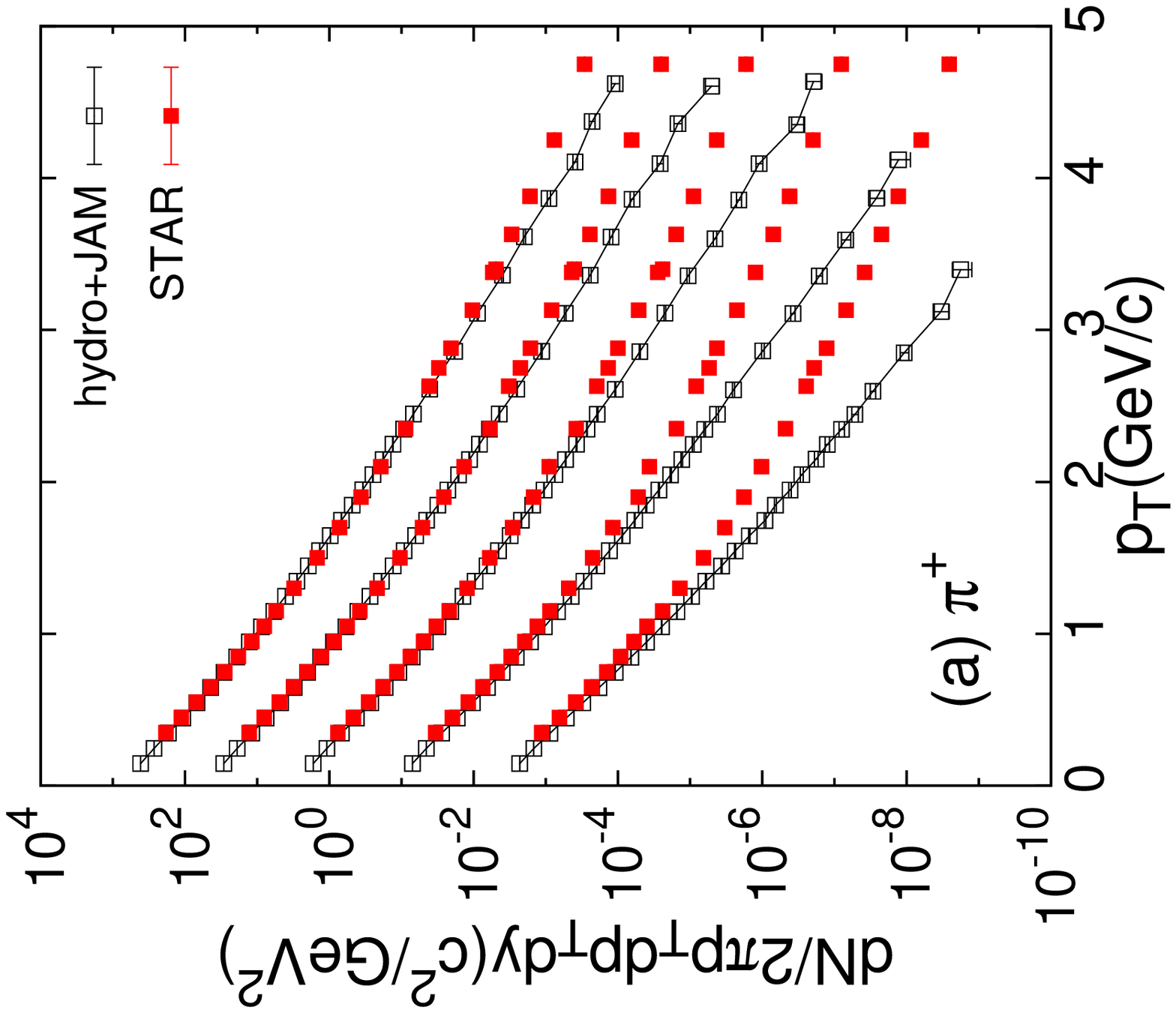}
\includegraphics[clip,angle=-90,width=0.46\textwidth]{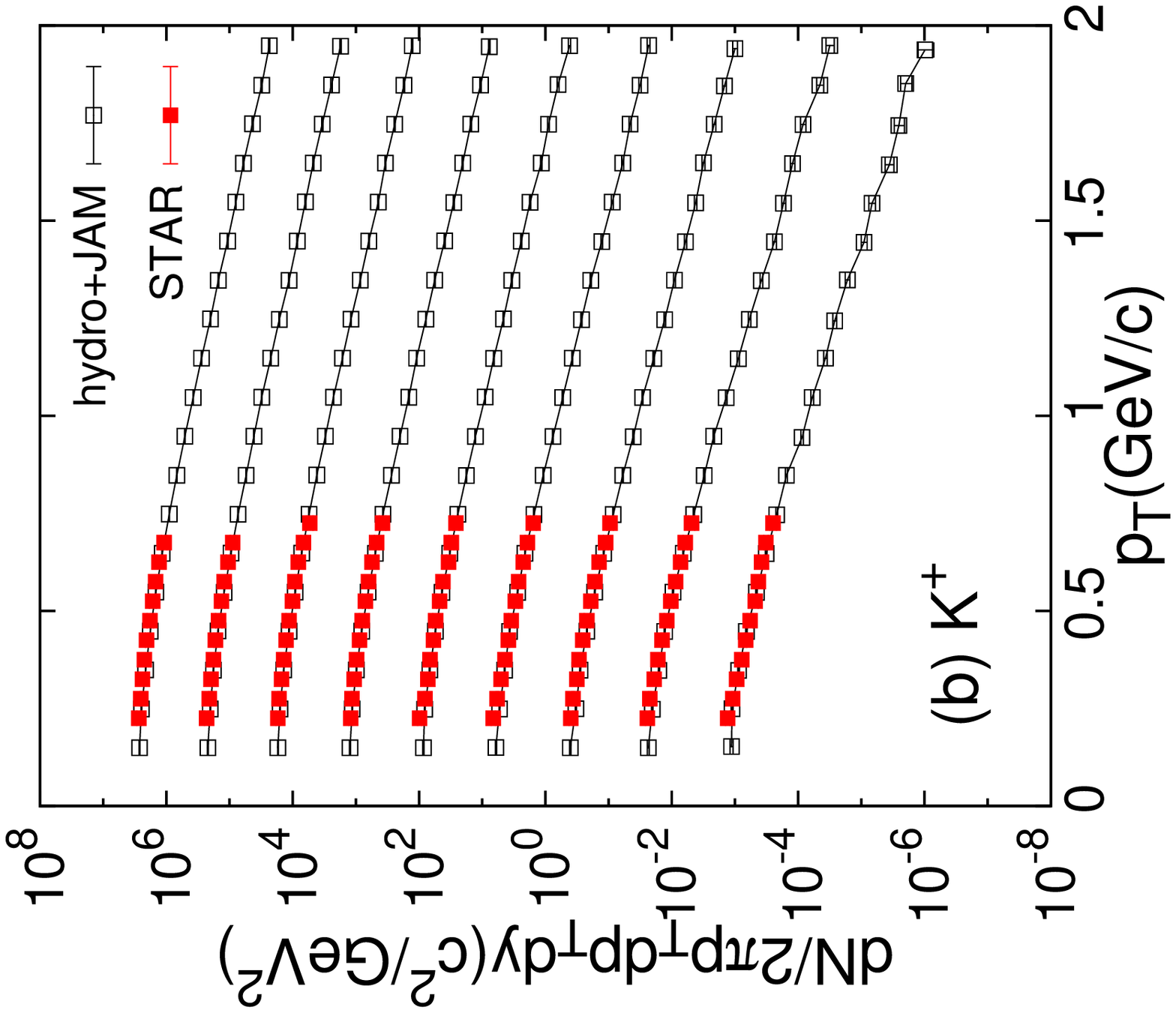}
\end{center}
\vspace{-5mm}
\caption{(Color Online) Transverse momentum distributions of (a)
  $\pi^{+}$ ($\left| \eta \right| < 0.5$) and (b) $K^{+}$ ($\left|
    \eta \right| < 0.1$) obtained from the integrated dynamical
  approach (open square) compared with data from the STAR
  Collaboration \cite{STAR:2006prl,STAR:2004prl} (filled square) for
  $\sNN =200\GeV$ Au+Au collisions.  From top to bottom, each spectrum
  shows the result of 0-12, 10-20, 20-40, 40-60, and 60-80\%
  centrality multiplied by $10^n$ with $n=0$ to $-4$ for pions, and
  0-5, 5-10, 10-20, 20-30, 30-40, 40-50, 50-60, 60-70, and 70-80\%
  centrality multiplied by $10^n$ with $n=5$ to $-3$ for kaons. }
\label{fig:dist_pi&k}
\end{figure*}

\begin{figure*}[tbp]
\begin{center}
\includegraphics[clip,angle=-90,width=0.46\textwidth]{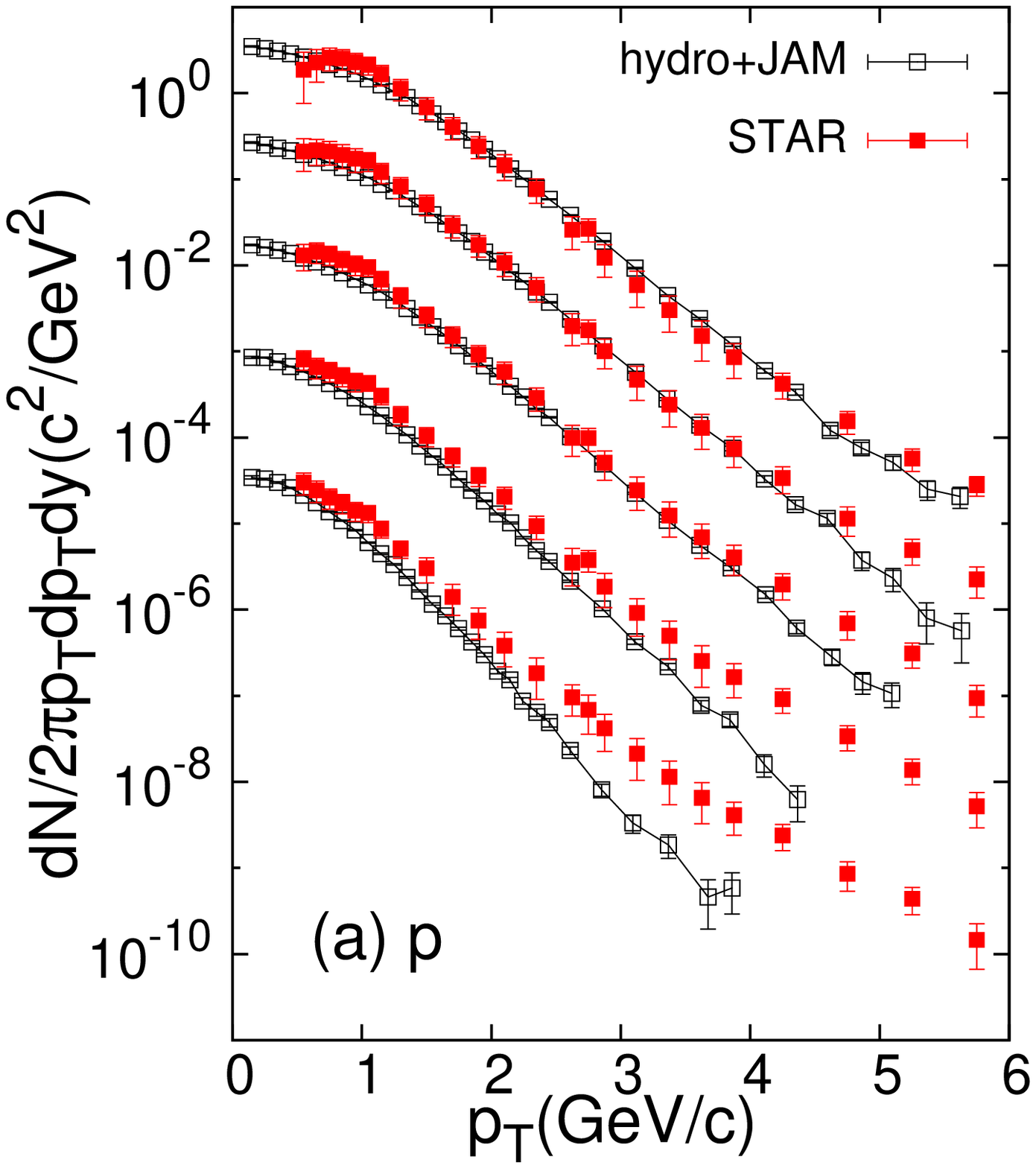}
\includegraphics[clip,angle=-90,width=0.46\textwidth]{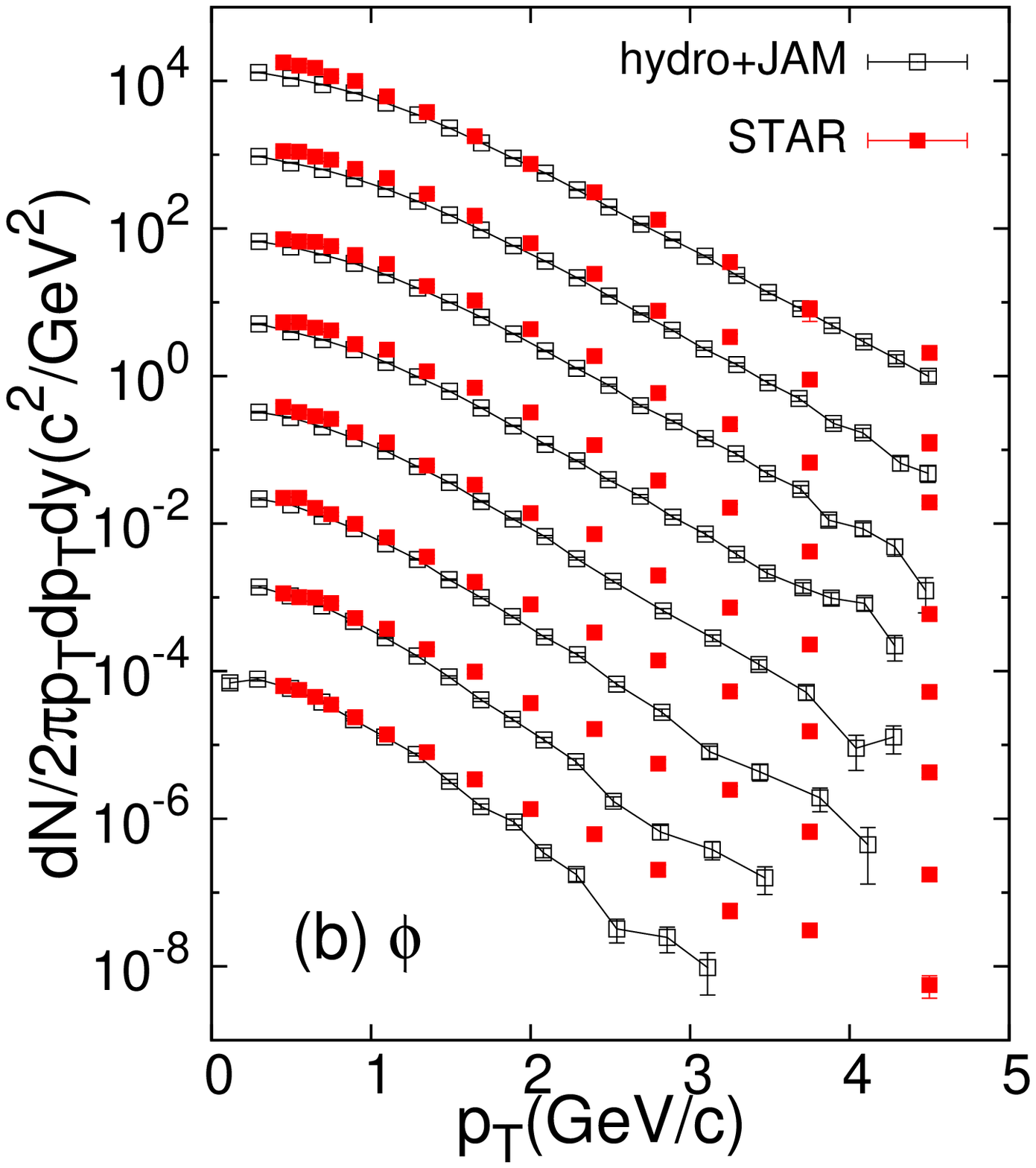}
\end{center}
\vspace{-5mm}
\caption{(Color Online) Transverse momentum distributions of (a)
  protons and (b) $\phi$ mesons in $\left| \eta \right| < 0.5$
  obtained from the integrated dynamical approach (open square)
  compared with data from the STAR Collaboration \cite{STAR:2006prl,
    STAR:2009prc} (filled square) for $\sNN =200\GeV$ Au+Au
  collisions.  From top to bottom, each spectrum shows the result of
  0-12, 10-20, 20-40, 40-60, and 60-80\% centrality multiplied by
  $10^n$ with $n=0$ to $-4$ for protons, and 0-10, 10-20, 20-30,
  30-40, 40-50, 50-60, 60-70, and 70-80\% centrality multiplied by
  $10^n$ with $n=4$ to $-3$ for $\phi$ mesons.  }
\label{fig:dist_pro&phi}
\end{figure*}

\begin{figure*}[tbp]
 \vspace{2mm}
\begin{center}
\includegraphics[clip,angle=-90,width=0.46\textwidth]{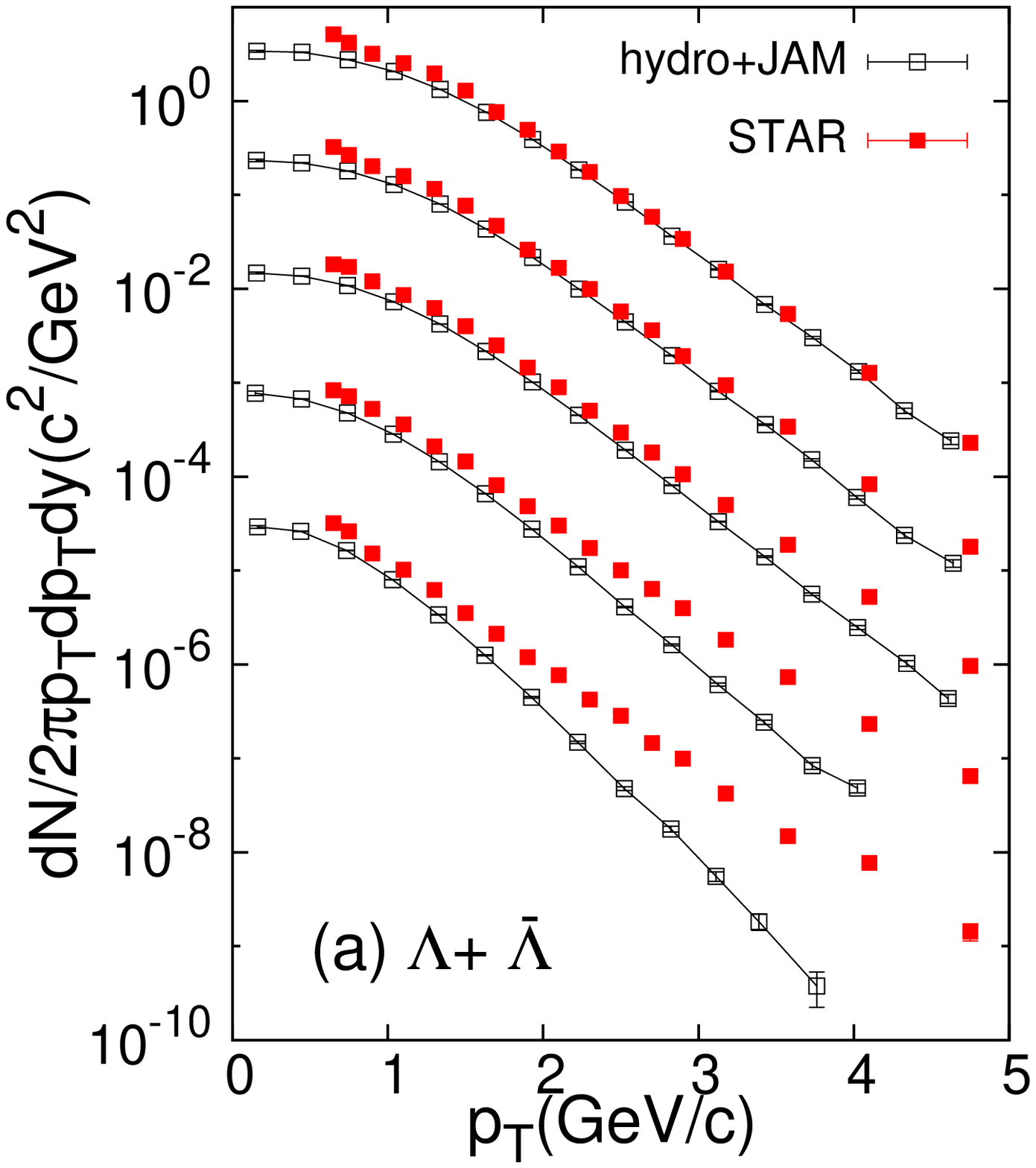}
\includegraphics[clip,angle=-90,width=0.46\textwidth]{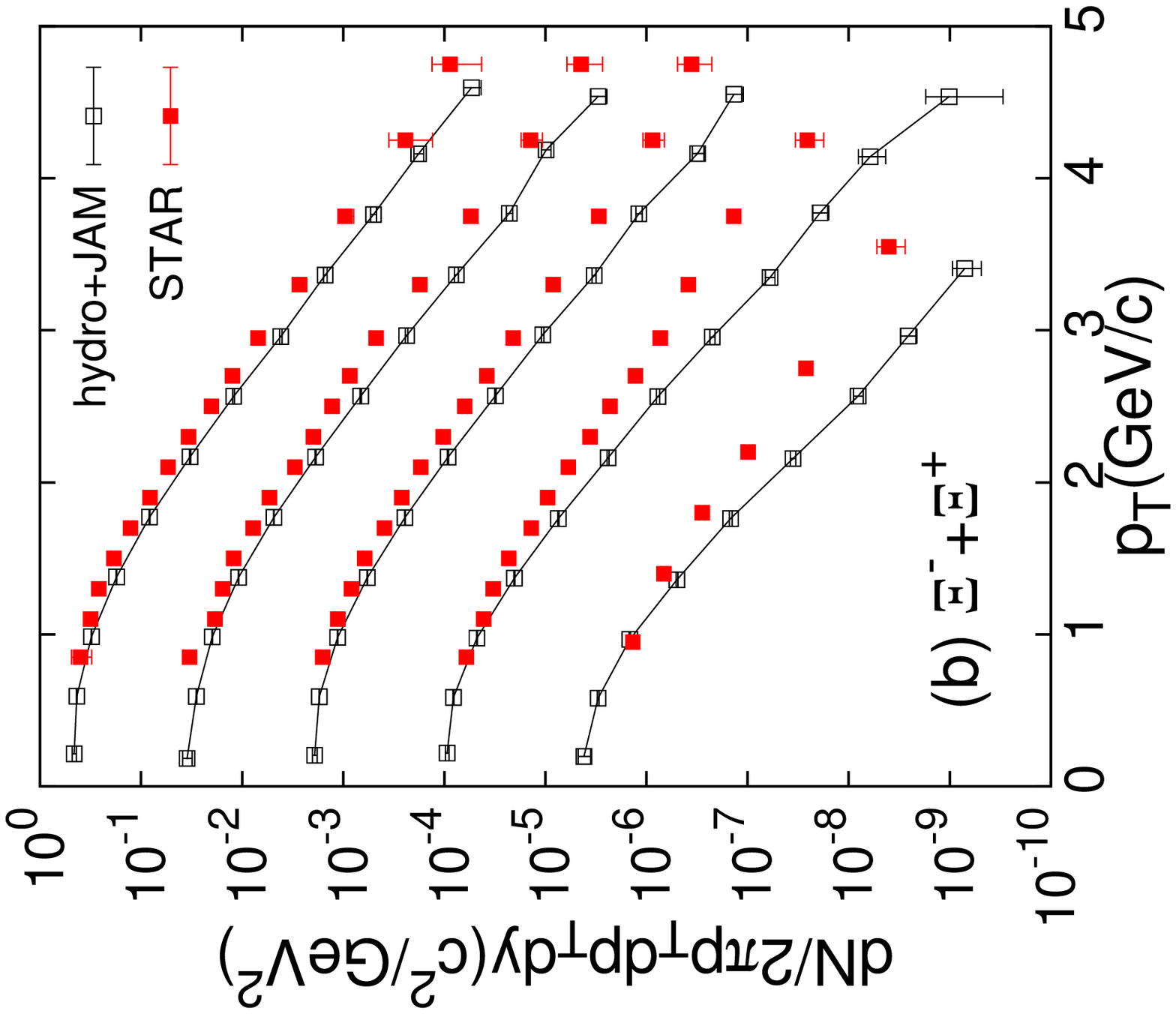}
\includegraphics[clip,angle=-90,width=0.46\textwidth]{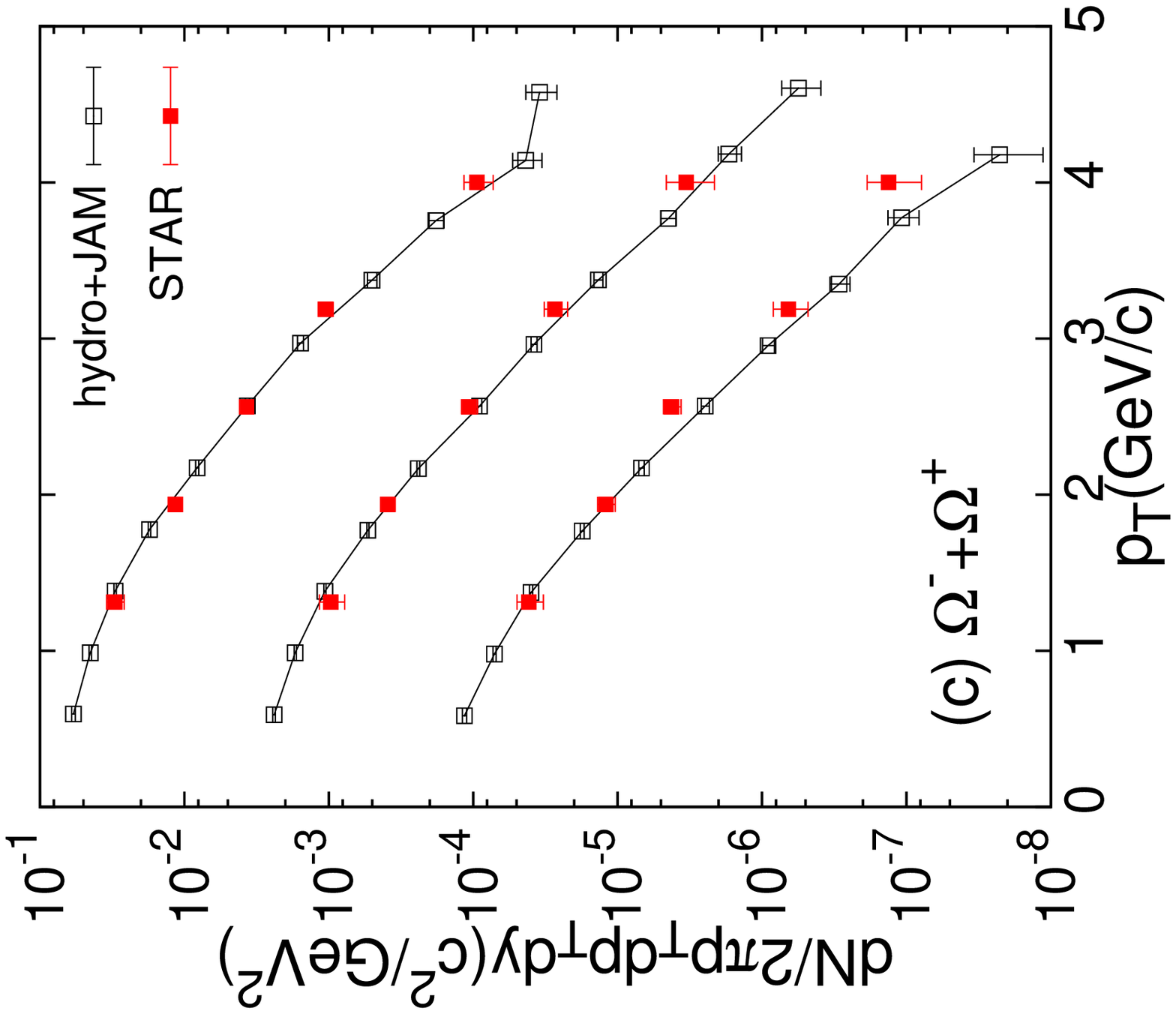}
\end{center}
\caption{(Color Online) Transverse momentum distributions of (a)
  $\Lambda \ (\left| \eta \right| < 1.0)$, (b) $\Xi \ (\left| \eta
  \right| < 0.75)$, and (c) $\Omega \ (\left| \eta \right| < 0.75)$
  obtained from the integrated dynamical approach (open square)
  compared with data from the STAR Collaboration \cite{STAR:2007prlb}
  (filled square) for $\sNN =200\GeV$ Au+Au collisions.  From top to
  bottom, each spectrum shows data of 0-5, 10-20, 20-40, 40-60, and
  60-80\% centrality multiplied by $10^n$ with $n=0$ to $-4$ for
  $\Lambda$ and $\Xi$ and 0-5, 20-40, and 40-60\% centrality
  multiplied by $10^n$ with $n=0$ to $-2$ for $\Omega$.  Note that the
  $\pT$ distributions of $\Lambda$ include the contribution from
  $\Sigma^0$ decay, see the text for details.  }
\label{fig:dist_hyperon}
\end{figure*}

In Figs.~\ref{fig:dist_pi&k}, \ref{fig:dist_pro&phi}, and 
\ref{fig:dist_hyperon}, 
we show $\pT$ distributions of identified hadrons around midrapidity 
in Au+Au collisions at $\sNN = 200 \GeV$. 
For clarity, the results and the experimental data points are scaled up or down 
by a common factor.

Figure~\ref{fig:dist_pi&k} shows the $p_{T}$ spectra
of (a) positive pions ($\left| \eta \right| < 0.5$) 
and (b) positive kaons ($\left| \eta \right| < 0.1$)
compared with the STAR data ($\left| y \right| < 0.5$ for 
pions~\cite{STAR:2006prl} and $\left| y \right| < 0.1$ for
kaons~\cite{STAR:2004prl}). The Jacobian between $\eta$ and $y$ is taken 
into account to obtain the invariant $\pT$ spectra in our results.
Even if we tuned the initial conditions and the switching temperature
in the model using pion, kaon, and proton + anti-proton
$p_{T}$ spectra from the PHENIX Collaboration in our previous study,
the results are overall in good agreement with the STAR data too.

In particular, the model reproduces the pion data up to $\pT \sim 3$
GeV/$c$ in central (0-12\%) collisions. Above this, our result
gradually deviates from the data due to appearance of (semi-)hard
components such as recombination and jet fragmentation. The more
peripheral the collision, the lower the $\pT$ where our result begins
to deviate from the data.  Due to the limited $\pT$ range of the data,
the same behavior is not seen in the kaon spectra. If one looks at the
pion spectrum in peripheral collisions (60-80\% centrality) carefully,
one sees that experimental data are systematically larger than the
results. This is because in peripheral collisions there is a small
discrepancy between PHENIX~\cite{Adler:2003cb} and
STAR~\cite{STAR:2006prl} pion yields. Nevertheless we did not further
tune the parameters to reproduce the STAR data because we want to keep
the same framework as in our previous study~\cite{Hirano:2013ppnp}.

 In Fig.~\ref{fig:dist_pro&phi}, we compare the $p_{T}$ spectra 
in $\left| \eta \right| < 0.5$ 
for (a) protons and (b) $\phi$ mesons 
with the STAR data \cite{STAR:2006prl, STAR:2009prc}. 
Due to limited statistics, we do not show some data points of the proton spectra
in high $p_{T}$ regions (above $\pT \sim 4$ GeV/$c$) in 20-40, 40-60,
and 60-80\% centrality.
 The proton spectra from STAR are corrected 
 for the $\Lambda$ and $\Sigma^{+}$ feed down.
Weak decays do not occur in the default setting in JAM, so 
we are able to compare our results directly with the STAR corrected data.
Since we neglect baryon chemical potential in our model,
our results for protons are slightly below the experimental data
at some $p_{T}$ range.
Nevertheless, overall slopes 
are in good agreement with the data
below $\pT \sim 3$ GeV/$c$.
On the other hand, similar to pions,
deviation between the experimental 
data and the results gradually increases with $p_{T}$, 
and the more peripheral the collision, the lower the $p_T$ where this
deviation appears.
As mentioned in the previous section, we switch off the decays 
of $\phi$ mesons in the hadronic cascade to be
able to analyze $\phi$ meson spectra directly without
resorting to mass reconstruction from kaons.  
Tendencies seen in $\phi$ meson spectra are similar to those in proton spectra.

Figure~\ref{fig:dist_hyperon} shows the $p_{T}$ spectra of
strange baryons, 
 (a) $\Lambda +\bar{\Lambda} \ (\left| \eta \right| < 1.0)$, 
 (b) $\Xi +\bar{\Xi} \ (\left| \eta \right| < 0.75)$, and 
 (c) $\Omega +\bar{\Omega}  \ (\left| \eta \right| < 0.75)$, 
 compared with the STAR data~\cite{STAR:2007prlb}.
Similar to the proton case, the more peripheral the collisions, 
 the worse the agreement
 between experimental data and the results, especially 
above $p_{T} = 1.5$-2.0 GeV/$c$.
As expected the measured $\Lambda$ and $\Xi$ yields are larger than
the calculated yields. We had chosen the switching temperature
$T_{\mathrm{sw}}$ to reproduce the observed pion to kaon and pion to
proton ratios, which leads to a lower temperature than the statistical
model fit to all observed
hadrons~\cite{Cleymans:2004pp,Huovinen:2007xh}, and thus to lower
yields of hyperons. On the other hand
it has been argued that the statistical model \cite{Andronic:2008gu}
with the common chemical freezeout temperature
does not reproduce the proton and $\Lambda$ yields 
simultaneously at the LHC and RHIC energies \cite{Milano:2013sza}. 
It was claimed~\cite{Steinheimer:2012rd} 
that earlier switching from hydrodynamics to UrQMD
at energy density $\sim 840$ MeV/fm$^3$, which corresponds to
higher temperature than $T_{\mathrm{sw}}=155$ MeV,
 would be a key to resolve 
 the issue, but our results do not improve significantly if we use
higher switching temperature $T_{\mathrm{sw}}=165$ MeV.
It remains to be seen whether the origin of this discrepancy lies in a
different description of the proton-antiproton annihilations in JAM
and UrQMD cascades, or in different expansion dynamics in collisions
at RHIC and LHC energies.

Note  that the experimental yield of $\Lambda$ in STAR contains 
the feed down from electromagnetic $\Sigma^0$ decays 
($\Sigma^{0} \rightarrow \Lambda + \gamma$)~\cite{STAR:2002prl}.
Since this is the dominant mode of $\Sigma^{0}$ decay with the branching 
ratio $\sim 100\%$ and $c \tau = 2.22 \times 10^{-11}$ m~\cite{PDG:2014CPC},
the feed down correction of it is experimentally challenging. 
We simulate this decay in JAM and include this contribution in the
final $\Lambda$ spectra in Fig.~\ref{fig:dist_hyperon} (a).
Since the mass of $\Sigma^{0}$ is close to that of $\Lambda$, the
yield of the primordial $\Sigma^{0}$ is expected to be of the same
order as  the yield of the primordial $\Lambda$. The ratio of
$\Sigma^0$ to $\Lambda$ at switching temperature can be estimated as
\begin{equation}
\frac{n_{\Sigma^0}}{n_{\Lambda}} \approx \frac{\exp \left(-\frac{m_{\Sigma^0}}{T_{\mathrm{sw}}} \right)}{\exp \left(-\frac{m_{\Lambda}}{T_{\mathrm{sw}}} \right)} \sim 0.61.
\end{equation}
After correcting for the effects of resonance decay and rescatterings,
we find that the final particle ratio of $\Sigma^0$ to $\Lambda$ is
around 0.3.
We have included both sources of $\Lambda$ in our results, but in general this
  should be kept in mind when comparing theoretical results with the
  data.

Although the number of data points from STAR is limited for $\Omega$
baryons, their yields and slopes are consistent with our results. It
is worth noticing here that the recent lattice QCD calculations
suggest the existence of resonances in
general~\cite{Majumder:2010ik}\footnote{It is, however, unknown
whether the difference between the lattice QCD and hadron resonance
gas model trace anomalies indicates the inadequacy of the hadron
resonance gas model~\cite{Wiranata:2013oaa}, or the existence of so
far undiscovered resonance states~\cite{Majumder:2010ik}.}, or strange
baryon resonances in particular~\cite{Bazavov:2014prl}, which have not
been discovered yet.

Such resonances, whether they are of Hagedorn kind~\cite{Hagedorn:1965st}, or those predicted by quark models (see e.g.~Refs.~\cite{Loring:2001kx,Loring:2001ky}), would contribute to the yields of strange hadrons, and might somewhat change the dynamics of the late hadronic stage~\cite{Greiner:2004vm,Noronha-Hostler:2013rcw,Noronha-Hostler:2014usa}. Some of those resonances could be formed in the scattering of $\phi$ mesons or $\Omega$ baryons, but it is unknown whether such resonances would significantly affect the average scattering cross section of $\phi$ mesons and $\Omega$ baryons in temperatures below the switching temperature. Investigation of this possibility is beyond the scope of the present paper.

\subsection{Elliptic flow}
\label{subsec:v2}

\begin{figure*}[tbp]
 \vspace{2mm}
\begin{center}
\includegraphics[clip,angle=-90,width=0.32\textwidth]{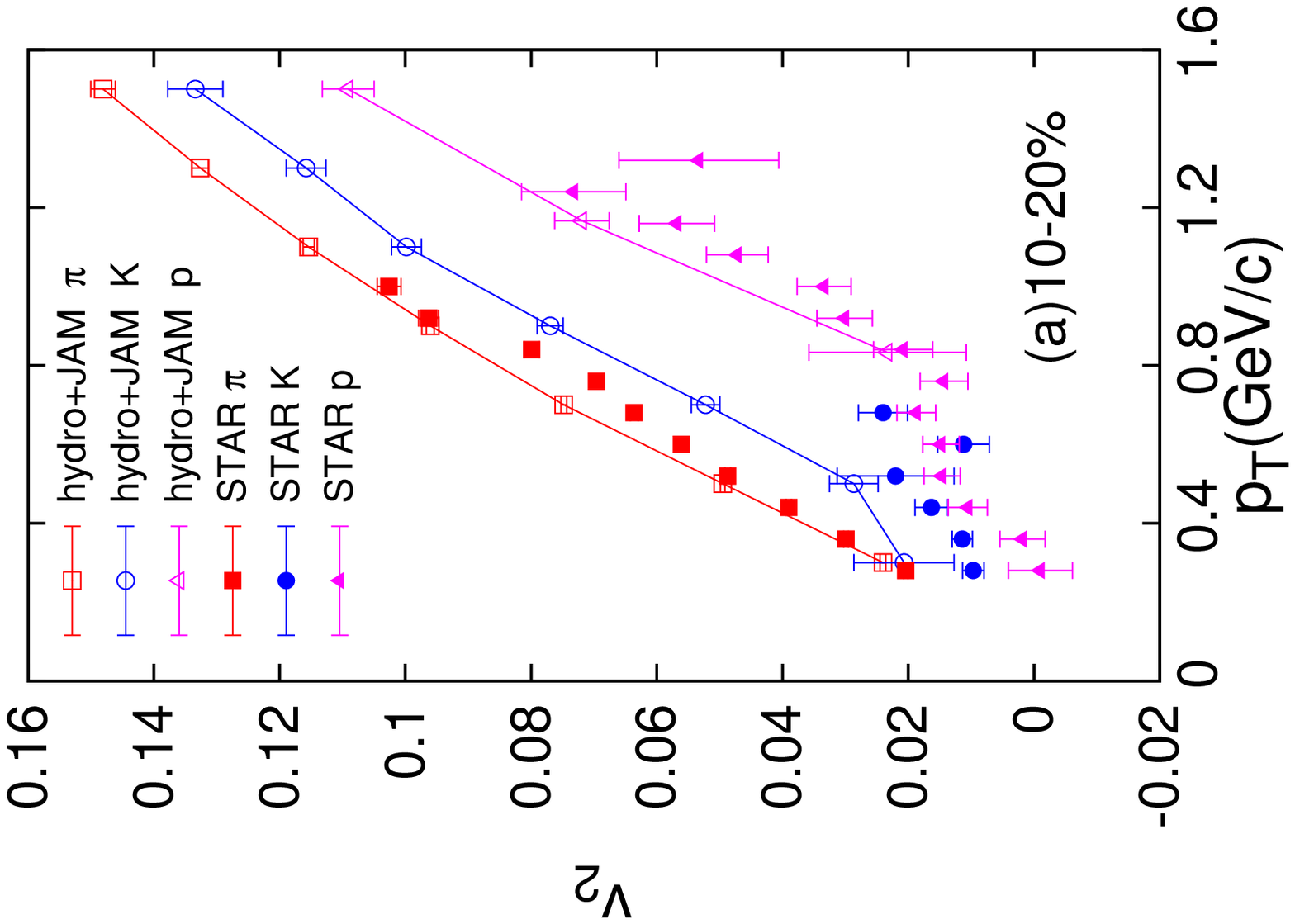}
\includegraphics[clip,angle=-90,width=0.32\textwidth]{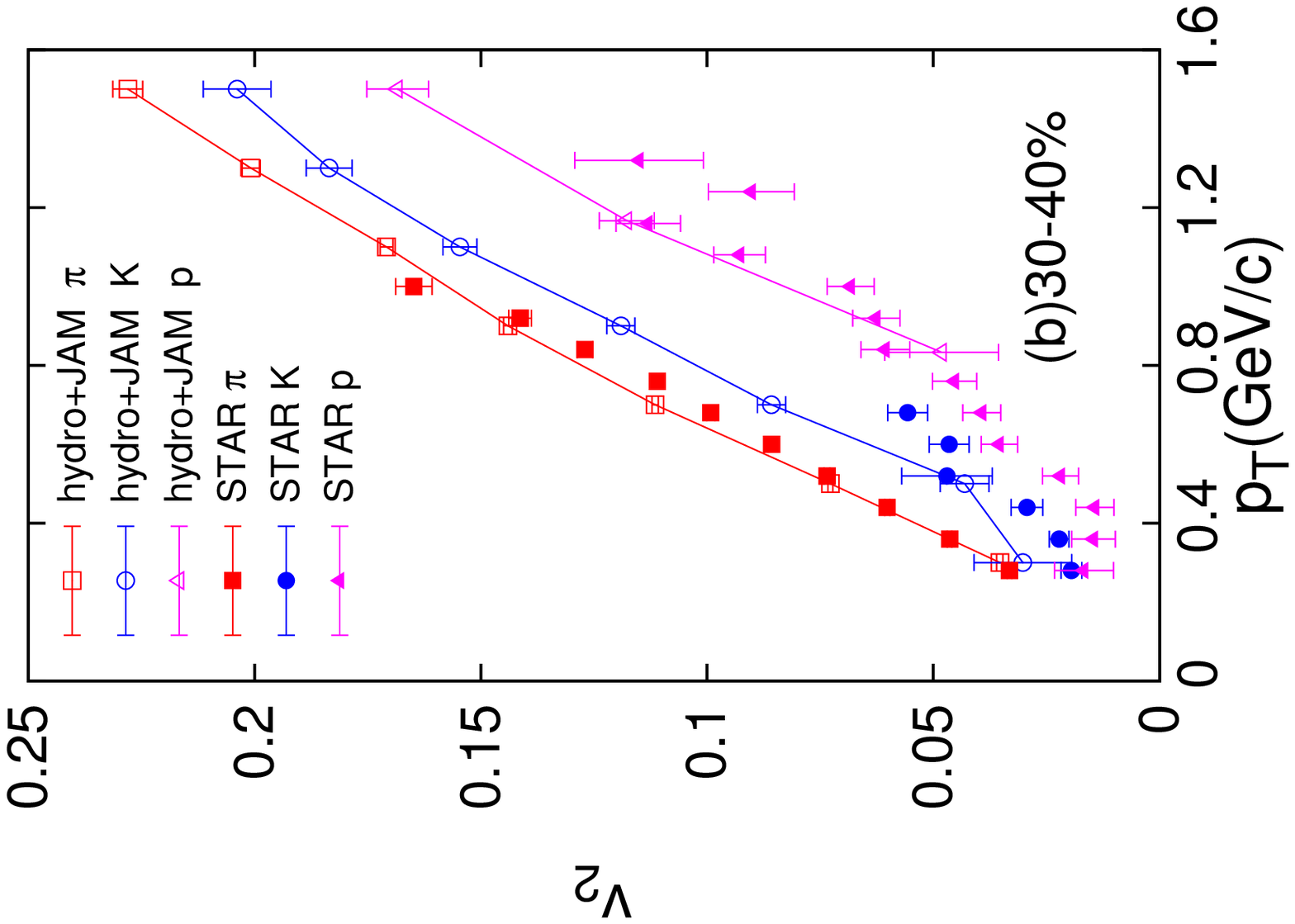}
\includegraphics[clip,angle=-90,width=0.32\textwidth]{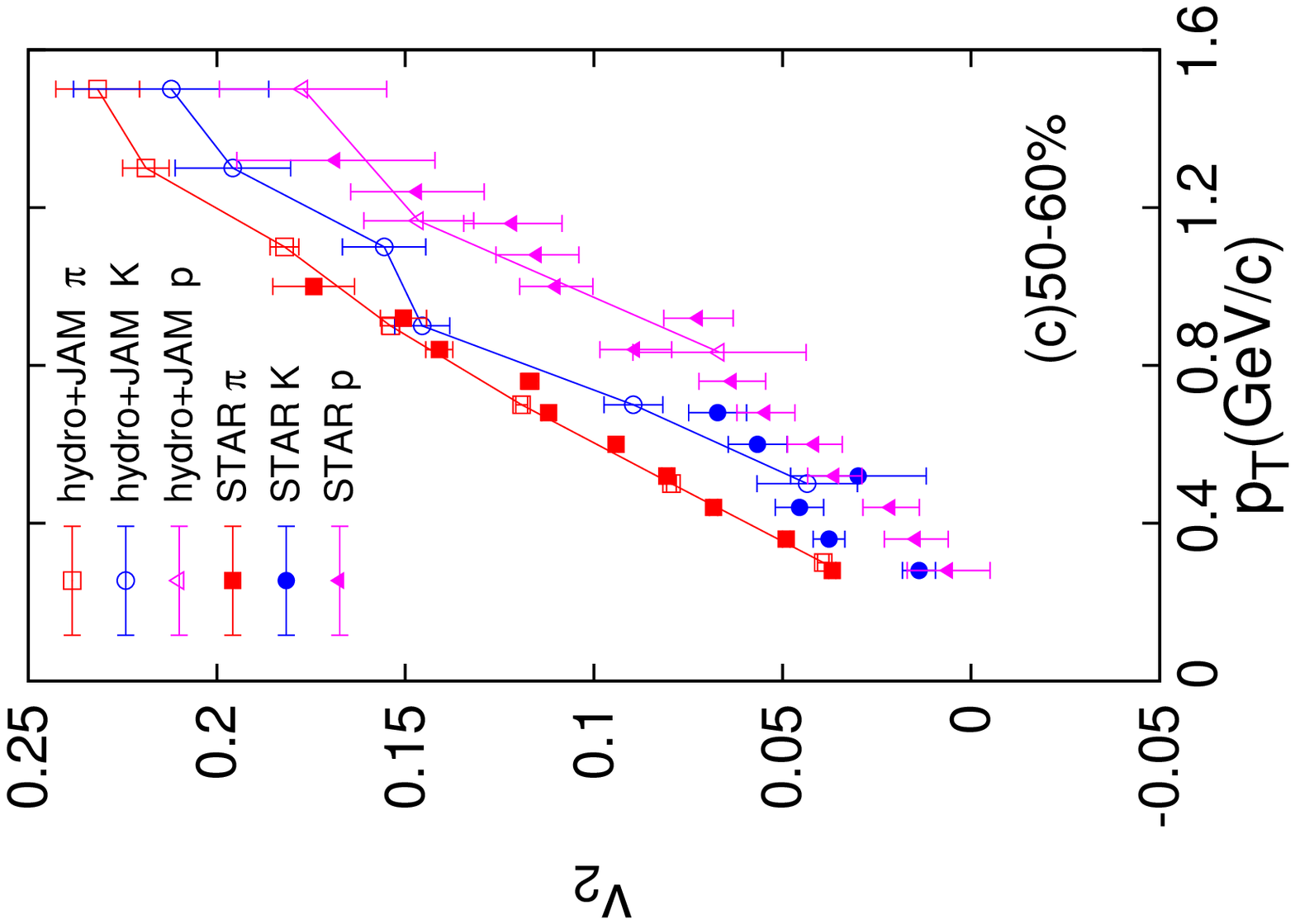}
\end{center}
\caption{(Color Online) Transverse momentum dependence of elliptic
  flow parameter $v_2\{2\}$ of pions (open square), kaons (open
  circle), and protons (open triangle) in $\left| \eta \right| < 1.0$
  obtained from the integrated dynamical approach compared with
  $v_2\{2 \}$ data from the STAR Collaboration \cite{STAR:2005prc}
  (filled symbol) for $\sNN=200\GeV$ Au+Au collisions. Centrality
  classes are (a) 10-20\%, (b) 30-40\%, and (c) 50-60\%. }
\label{fig:v2_piKp1}
\end{figure*}

In Fig.~\ref{fig:v2_piKp1} the two-particle cumulant $v_{2}(\pT)$ of
pions, kaons, and protons in the midrapidity region 
($\left| \eta \right| < 1.0$)
are compared with STAR $v_{2}\{ 2\}$ data \cite{STAR:2005prc}
at three centralities (10-20\%, 30-40\%, and 50-60\%).
For pions, we reasonably reproduce the experimental data for each
centrality class. Due to the limited number of events, it is hard to
obtain $v_{2}\{2\}$ with smaller statistical errors, and especially
difficult this would be for protons. Thus some points with large
error bars, in particular in low $\pT$ region, are not shown in these
figures for clarity. We also calculated $v_{2}$\{RP\} (not shown),
namely $v_{2}$ with respect to theoretically-known reaction plane, to
reduce the statistical errors and found that $v_{2}$\{RP\} of pions,
kaons, and protons from our model are in reasonable agreement with the
STAR data in mid-central collisions. However, $v_{2}$\{RP\} lacks
initial fluctuations of the event plane angle and, consequently, is
significantly smaller than the data in both central and peripheral
collisions. It is noted that, from the MC Glauber model analysis, the
fluctuation effect on eccentricity is large in central and peripheral
collisions, but almost negligible in mid-central collisions.  See,
\textit{e.g.}, Fig.~3 (a) in Ref.~\cite{Hirano:2009prc}.  At the
centralities shown in Fig.~{\ref{fig:v2_piKp1}} 
a clear mass ordering behavior among pions, kaons, and
protons, namely larger $v_{2}$ for smaller mass at fixed $\pT$, is
seen both in our results and in the STAR data.

\begin{figure*}[tbp]
\begin{center}
\includegraphics[clip,angle=-90,width=0.48\textwidth]{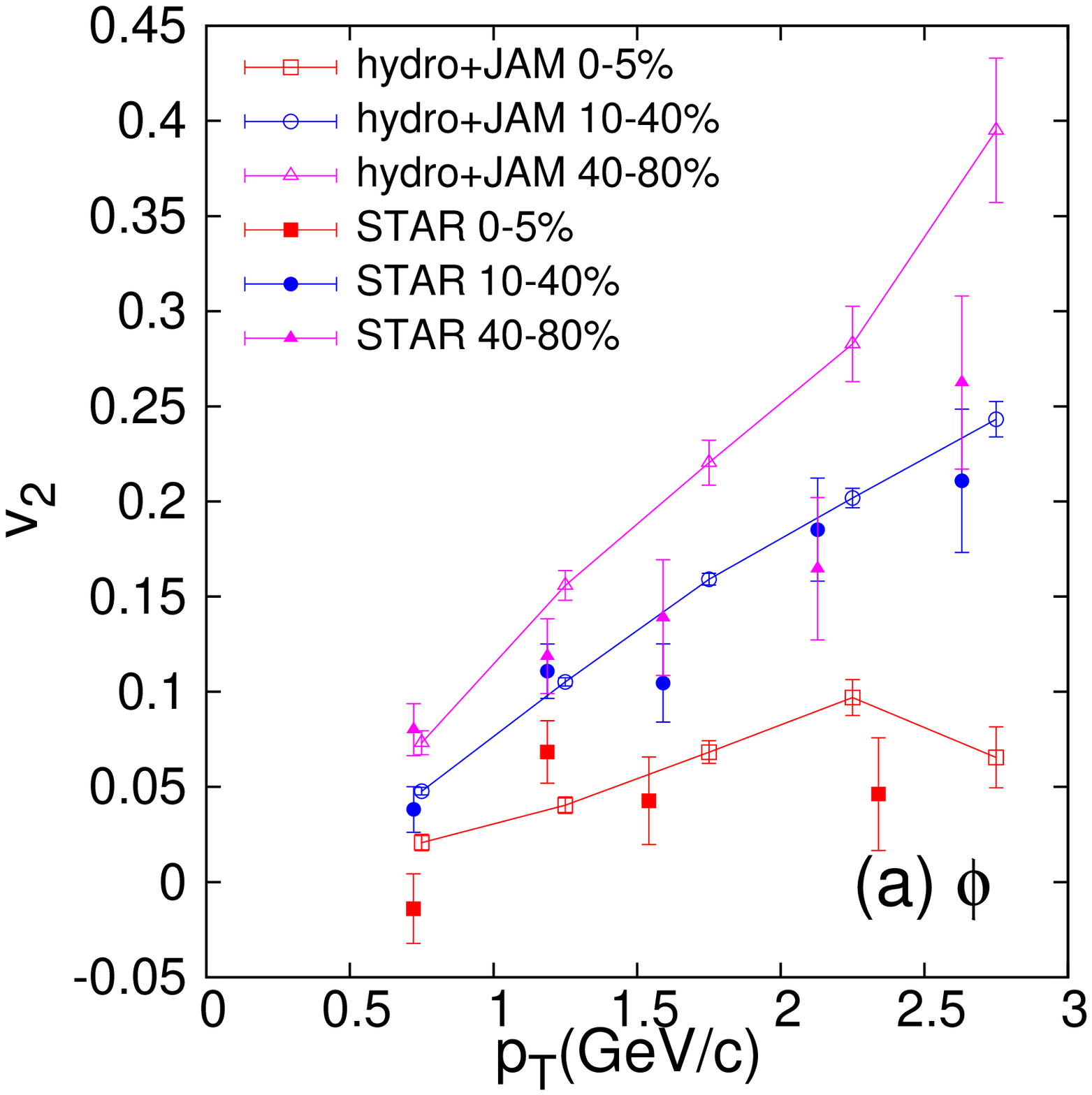}
\includegraphics[clip,angle=-90,width=0.48\textwidth]{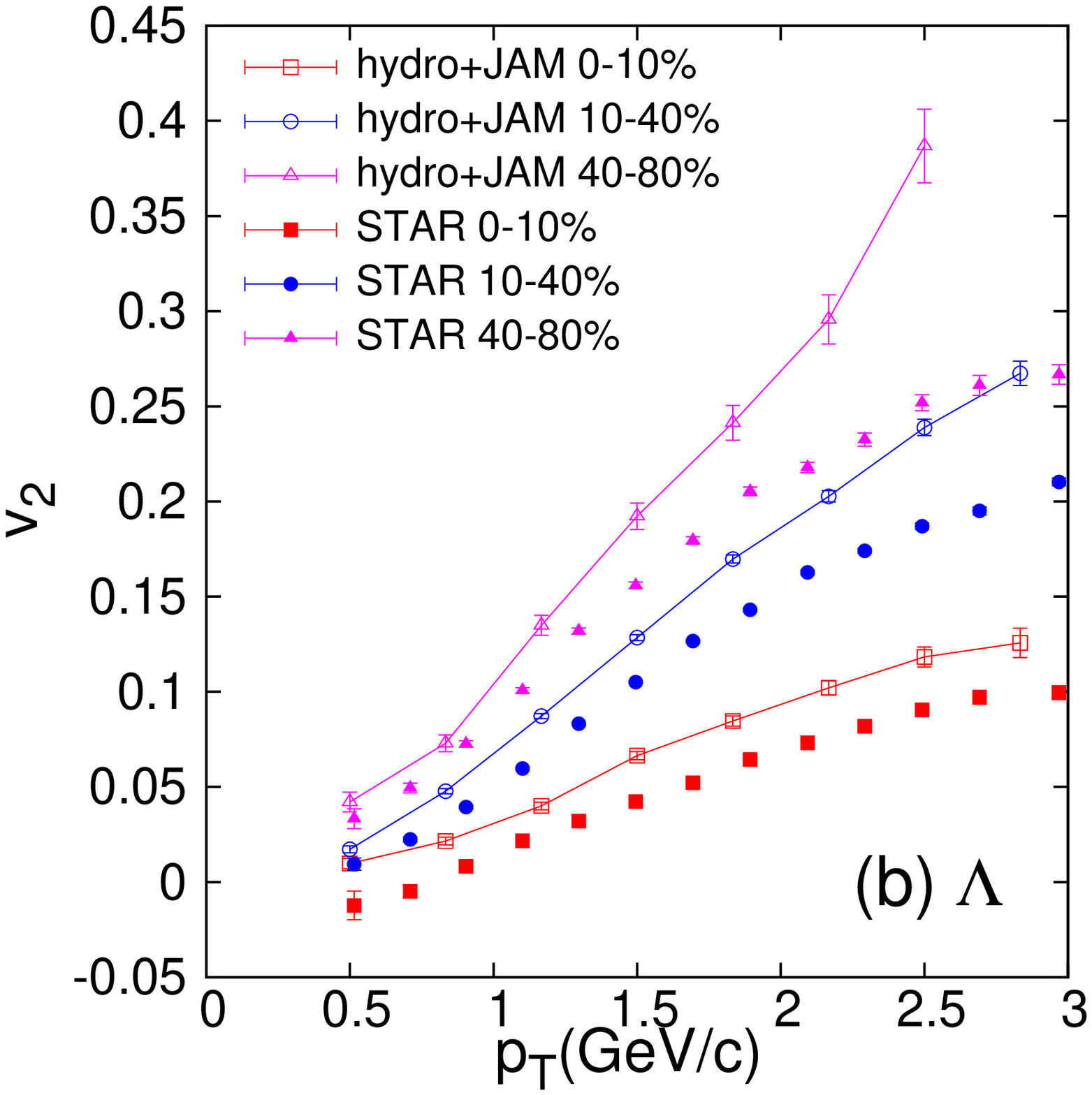}
\includegraphics[clip,angle=-90,width=0.48\textwidth]{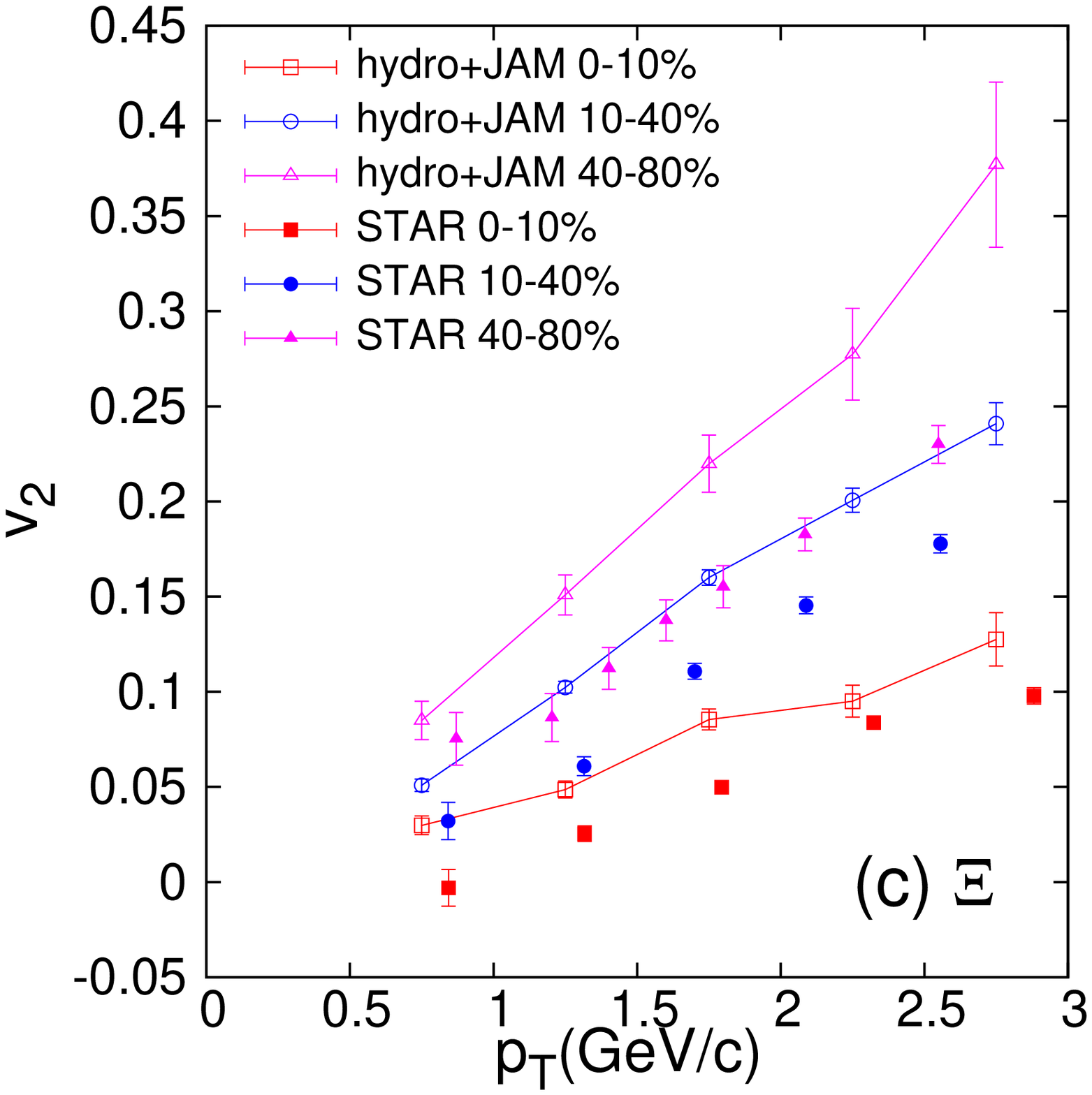}
\includegraphics[clip,angle=-90,width=0.48\textwidth]{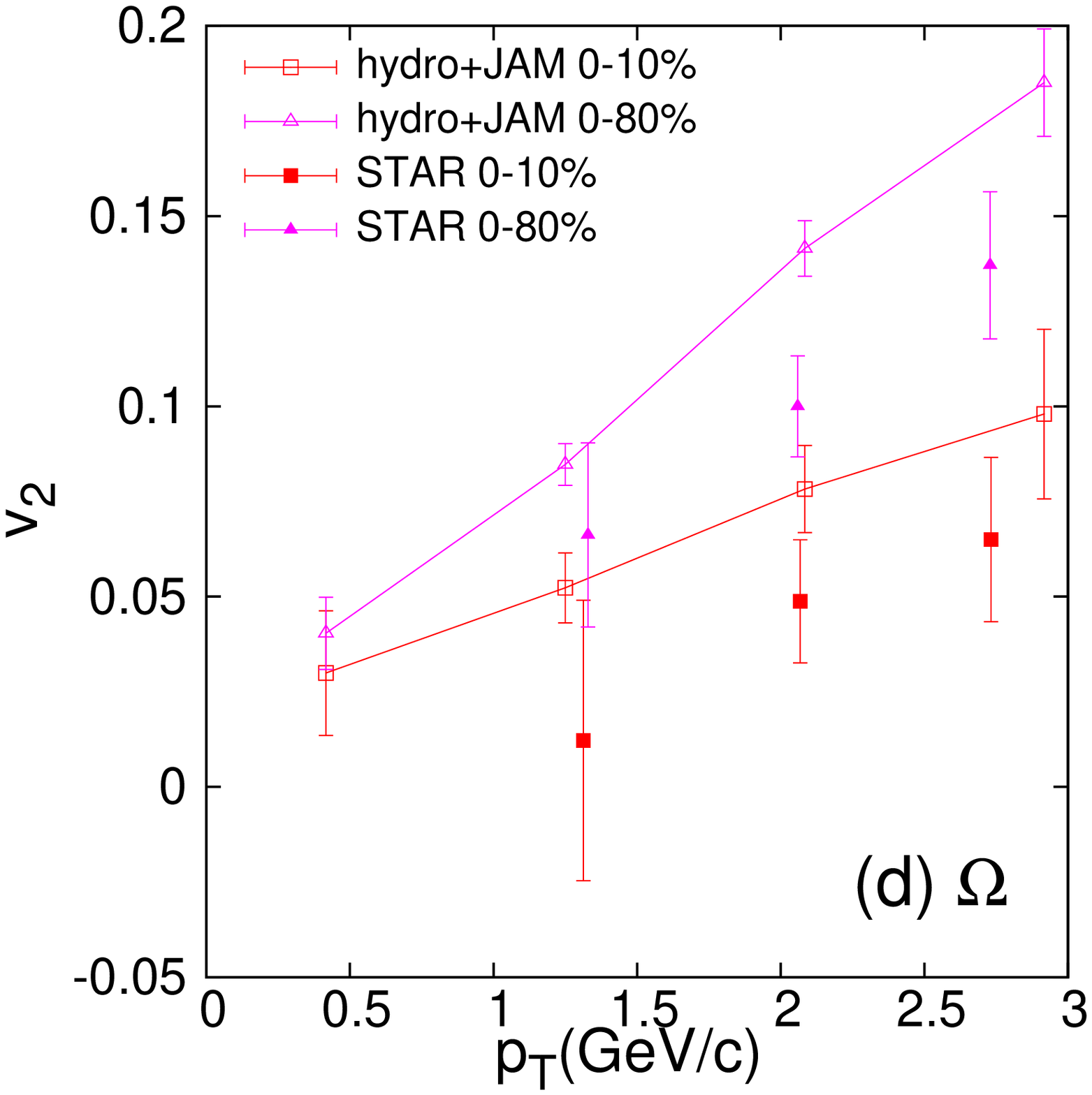}
\end{center}
\caption{(Color Online) Transverse momentum dependence of the elliptic
  flow parameter $v_{2}$\{EP\} of (a) $\phi$ mesons (0-5, 10-40, and
  40-80\% centrality), (b) $\Lambda$ baryons (0-10, 10-40, and 40-80\%
  centrality), (c) $\Xi$ baryons (0-10, 10-40, and 40-80\%
  centrality), and (d) $\Omega$ baryons (0-10 and 0-80\% centrality)
  in $\left| \eta \right| < 1.0$ obtained from the integrated
  dynamical approach (open symbol) compared with data from the
  STAR Collaboration \cite{STAR:2009prc,STAR:2008prc} (filled symbol)
  for $\sNN=200\GeV$ Au+Au collisions.  }
\label{fig:v2_hyperon}
\end{figure*}

In Fig.~\ref{fig:v2_hyperon}, 
we show the event plane $v_{2}(\pT)$ 
of $\phi$, $\Lambda$, $\Xi$, and $\Omega$ at some centrality classes
from the integrated dynamical approach.  
In comparison, STAR data from the event plane method $v_{2}$\{EP\} 
\cite{STAR:2009prc,STAR:2008prc} are also shown in the figure.
We calculate $v_{2}$\{EP\} by employing 
the $\eta$-subevent method using hadrons in $|\eta | < 1.0$ 
as done by the STAR Collaboration~\cite{STAR:2008prc},
\begin{equation}
v_{2}\{\mbox{EP}\} = \frac{1}{\mathcal{R}} \langle \cos [2 (\phi_\pm -\Psi_{2,\mp})] \rangle.
\end{equation}
The event plane angle $\Psi_{2,+}$ ($\Psi_{2,-}$) is defined for
the particles with positive (negative) pseudorapidity.
In the actual experimental analysis, an pseudorapidity  gap of 
$|\Delta \eta| = 0.075$ between two 
subevents was introduced to suppress the nonflow effects \cite{STAR:2008prc}.
We found, however, that introduction of the
 gap does not change final results within error bars 
in our analysis.
The resolution factor in this case
is calculated as $\mathcal{R} = \sqrt{\langle \cos [2(\Psi_{2,+}-\Psi_{2,-})] \rangle }$.
In calculating the $Q$ vector,
\begin{equation}
Q = \sum_j w_j \exp(2i\phi_j),
\end{equation}
needed for the event plane angle 
\begin{equation}
\Psi_{2,\pm} = \frac{1}{2}\tan^{-1}\frac{\Im Q}{\Re Q},
\end{equation}
a weight factor $w_j = \mbox{min}\{p_{T,j}, 2 \mbox{ GeV}/c\}$
was also introduced \cite{STAR:2008prc}.
However, within statistical errors the results are similar to those
obtained using $w_j=1$. Therefore, we employ $w_j=1$ in all the
results shown in Fig.~\ref{fig:v2_hyperon}.
For $\Lambda$ and $\Xi$ our results are systematically larger than the
STAR data. For $\phi$ and $\Omega$ our results show similar behavior,
but since the data have large errors, no firm conclusion can be made.

Our results of elliptic flow parameters for identified hadrons
 are systematically larger than the STAR data,
which suggests that there is room for finite, but perhaps small, viscosity
in the fluid-dynamical stage.
However, we do not go into details about the QGP viscosity since the
purpose of the present study is to investigate the effects of
rescattering in the late hadronic stage, which are not affected by
the QGP viscosity nor by the initial state of the system.

\section{Hadronic rescattering effects}
\label{sec:RescatteringEffects}

In this section, 
we investigate hadronic rescattering effects
on final observables,
in particular, 
how much final observables reflect the properties of the system 
at particlization in the integrated dynamical approach.
If some observables do not change much during the late kinetic stage,
these observables can be utilized as ``penetrating'' probes. Namely,
their distributions reflect the properties of the system immediately
after hadronization, and are not contaminated by later hadronic
rescatterings.
To quantify the effects of hadronic rescattering
we perform simulations of nuclear collisions with
the following three options in the hadronic cascade calculations:
(I) full dynamical evolution (default),
(II) deactivating hadronic rescatterings, and
(III) deactivating both hadronic rescatterings and resonance decays.
To simplify the calculations we 
calculate elliptic flow parameters with respect to the
theoretically-known reaction plane.

\begin{figure*}[tbp]
\begin{center}
\includegraphics[bb=50 50 554 554,angle=-90,width=0.48\textwidth]{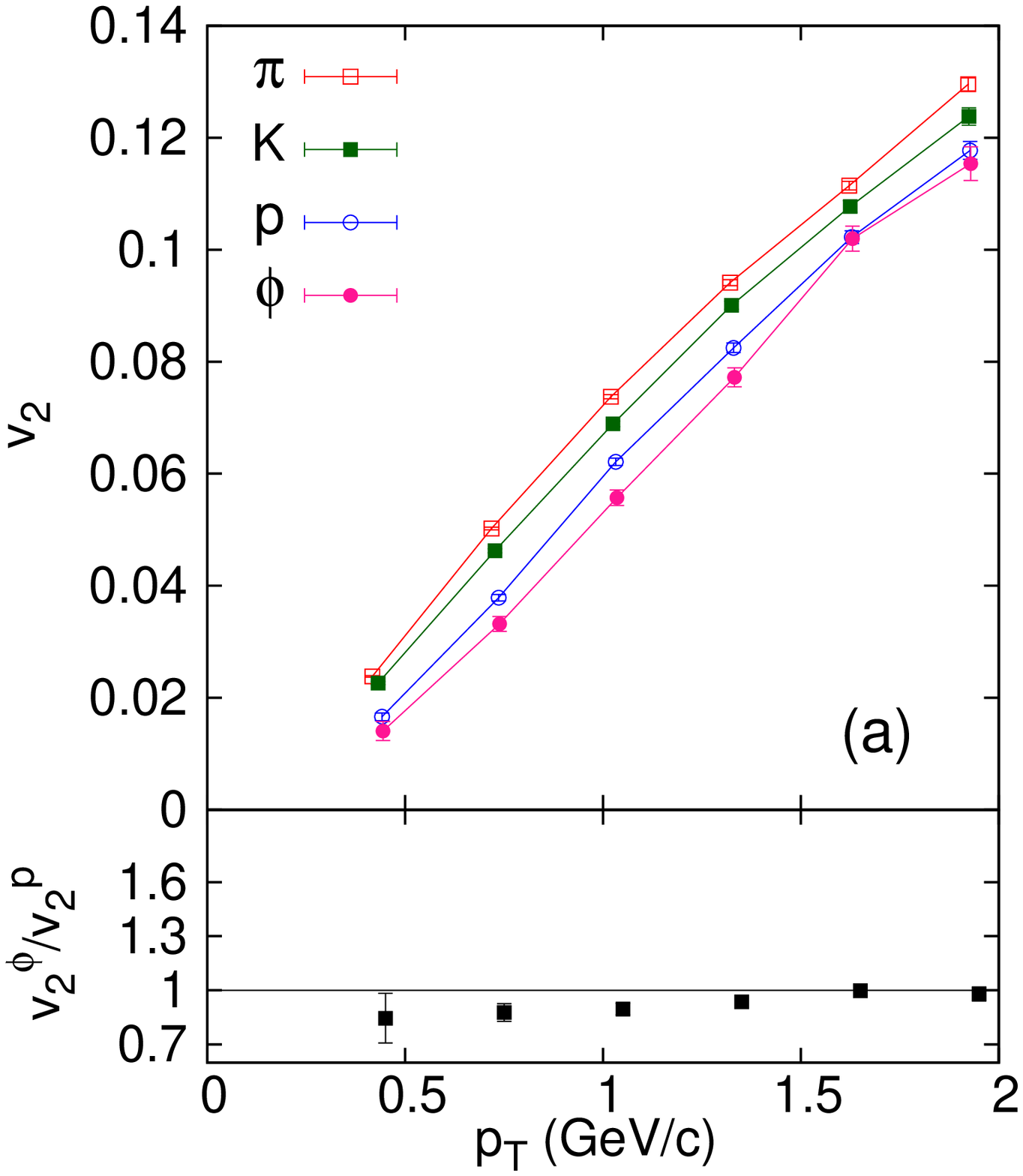}
\includegraphics[bb=50 50 554 554,angle=-90,width=0.48\textwidth]{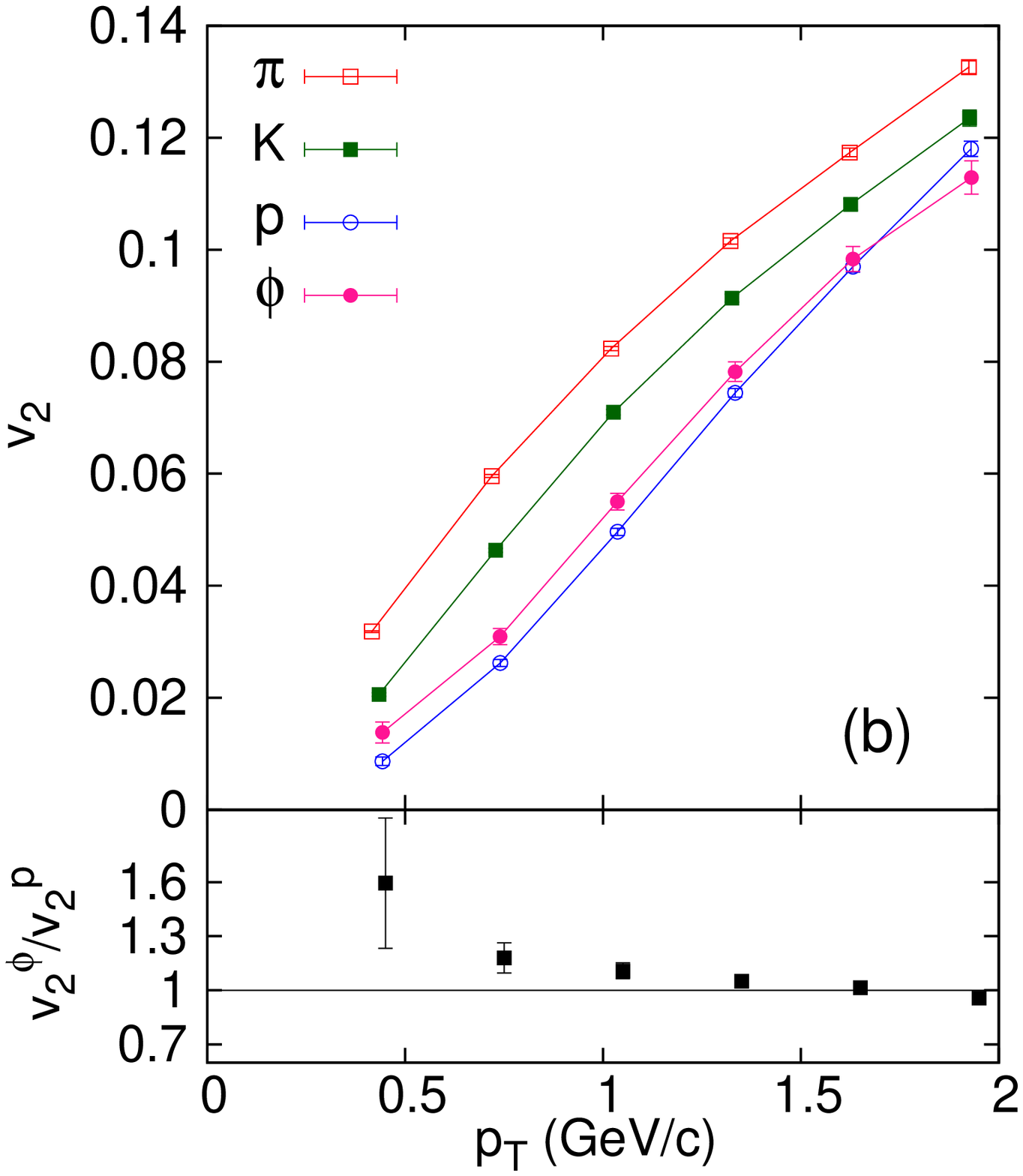}
\end{center}
\caption{(Color Online) Transverse momentum dependence of elliptic
  flow parameter $v_2$ for pions (open square), kaons (filled square),
  protons (open circle), and $\phi$ mesons (filled circle) in 
  $\left| \eta \right| < 1.0$ obtained from the integrated dynamical
  approach (a) without hadronic rescattering and (b) with hadronic
  rescattering in minimum bias $\sNN=200\GeV$ Au+Au collisions. The
  lower panels of the plots show the ratio of $v_{2}^{\phi}$ to $v_{2}^{p}$.}
\label{fig:v2_noscat_scat}
\end{figure*}

In Figs.~\ref{fig:v2_noscat_scat} (a) and (b),
we show $v_{2}(\pT)$ of pions, kaons, protons, and $\phi$ mesons 
in minimum bias Au+Au collisions at $\sNN =200\GeV$ 
from the integrated dynamical approach.
Figure~\ref{fig:v2_noscat_scat} (a) represents the results
 without hadronic rescatterings (option (II)).
These results exhibit the mass ordering behavior, namely  
${v_{2}^{\pi}(\pT)} > {v_{2}^{K}(\pT)} > {v_{2}^{p}(\pT)} > {v_{2}^{\phi}(\pT)}$ 
for $m_{\pi} < m_{K} < m_{p} < m_{\phi}$,
in the low $p_{T}$ region.
In general the mass ordering results from collective flow in which all
the components of the fluid flow at a common fluid
velocity~\cite{Huovinen:2001cy}.
On the other hand, Fig.~\ref{fig:v2_noscat_scat} (b) shows a
violation of mass ordering 
between protons and $\phi$ mesons below $p_{T} \sim 1.5$ GeV/$c$:
$v_{2}^{p}< v_{2}^{\phi}$ even though $m_{p}< m_{\phi}$.
If one compares these two plots,
${v_{2}^{p}(\pT)}$ decreases below $\sim$1.5 GeV/$c$
during the hadronic rescattering stage. 
On the other hand, ${v_{2}^{\phi}(\pT)}$ stays almost unchanged 
during the hadronic rescattering stage.
As a consequence, the order is inversed between protons and $\phi$
mesons. To see this behavior more clearly, we also show the ratio
$v_{2}^{\phi}/v_{2}^{p}$. Below 1 GeV/$c$, this ratio becomes clearly
positive indicating the violation of mass ordering.

The violation of mass ordering was predicted in Ref.~\cite{Hirano:2008prc}
and observed recently by the STAR Collaboration \cite{Nasim:2013npa}.
It should be noted that a larger switching temperature
($T_{\mathrm{sw}}=169$ MeV)
and a first order phase transition
in the EoS were employed in the calculations of Ref.~\cite{Hirano:2008prc}.
We confirm in the present study that the violation of
mass ordering is also observed when a more realistic
EoS from lattice QCD 
and a lower switching temperature ($T_{\mathrm{sw}}=155$ MeV) are used.

We can interpret  the violation of mass ordering 
as a result of evolution of mean $p_{T}$ and $p_{T}$-averaged $v_{2}$
during the late transport stage. 
The slope of the $p_{T}$-differential $v_{2}$
can be approximated 
by the ratio of $p_T$-averaged $v_{2}$ to mean $p_{T}$ \cite{Hirano:2005wx},
\begin{equation}
\label{eq:v2slope}
\frac{dv_{2}(\pT)}{ d\pT} \approx \frac{v_{2}}{\langle \pT \rangle}.
\end{equation}
If $v_{2}(p_{T})$ is linearly proportional to $p_{T}$, 
it is easy to show that Eq.~(\ref{eq:v2slope}) holds exactly. 
Thus a change of slope in $v_{2}(p_{T})$ due to hadronic rescatterings
could be attributed to changes of $v_{2}$ and/or $\langle \pT \rangle$,
which provides us with more intuitive understanding.

\begin{figure*}[tbp]
\begin{center}
\includegraphics[clip,angle=-90,width=0.48\textwidth]{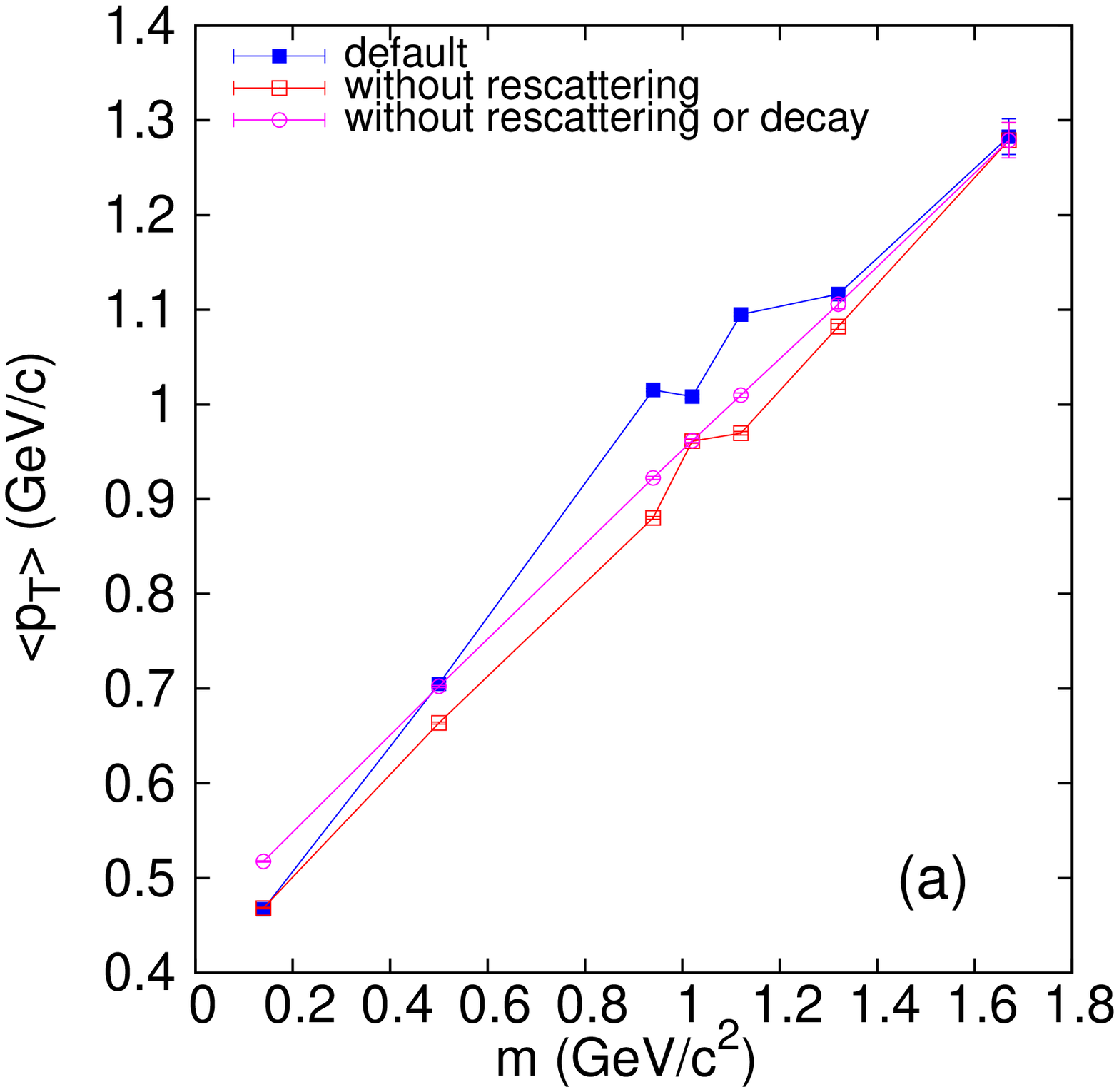}
\includegraphics[clip,angle=-90,width=0.48\textwidth]{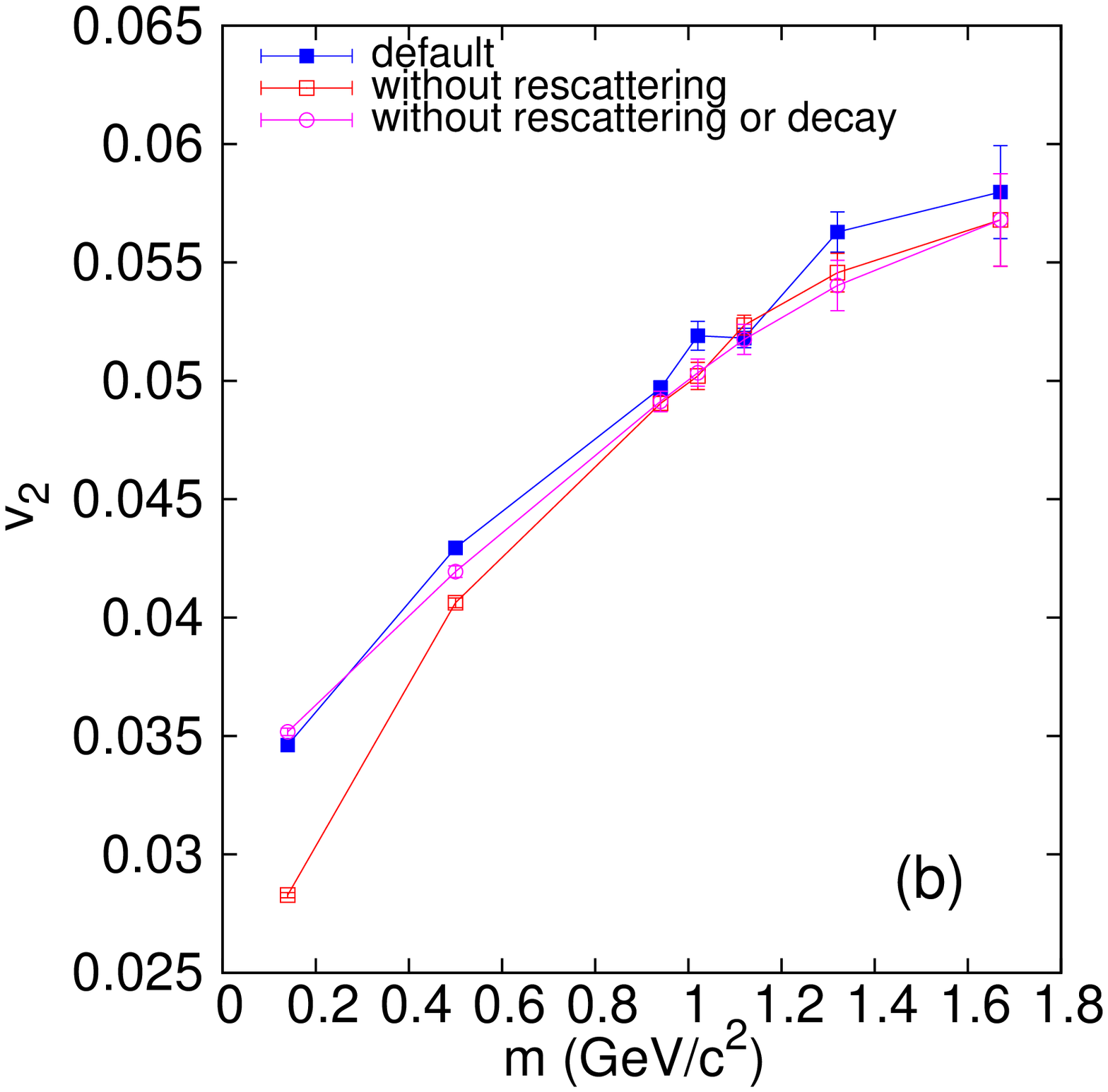}
\end{center}
\caption{(Color Online) (a) Mean transverse momentum $\langle \pT \rangle$ and
  (b) $\pT$-averaged $v_{2}$ in minimum bias $\sNN=200\GeV$
  Au+Au collisions around midrapidity ($\left| y \right| < 1.0$)
  obtained by the hybrid model as a function of the hadron mass.  Each
  point represents the result for $\pi$, $K$, $p$, $\phi$, $\Lambda$,
  $\Xi$, and $\Omega$ from left to right. Three lines show the results
  with different options respectively: (I) full calculations with both
  resonance decays and hadronic rescsatterings (filled square), (II)
  calculations with resonance decays only (open square), and (III)
  calculations without resonance decays or hadronic rescatterings
  (open circle). Lines are to guide the eye.}
\label{fig:pt&v2_versus_mass}
\end{figure*}

We next calculate $\langle \pT \rangle$ 
and $\pT$-averaged $v_2$ for each identified hadron 
in $|y|<1$
in minimum bias Au+Au collisions at $\sNN =200\GeV$ 
with the three different options mentioned above.
We show  $\langle \pT \rangle$ and $v_2$ as functions
of hadron mass in Figs.~\ref{fig:pt&v2_versus_mass} (a) and (b), respectively.
Default results using the option (I) are shown with closed squares. 
To see the effect of hadronic rescatterings, 
we use the option (II)  in the hadronic cascade calculation.
We also use the option (III) to obtain the results excluding the
contributions of both hadronic rescatterings and resonance decays.
In the default option (I) in which both hadronic rescatterings and resonance decays
are included,  $\langle \pT \rangle$  is not a linear function of the hadron mass.
By switching off hadronic rescatterings by employing the option (II),
we are able to extract the hadronic rescattering effects on evolution of mean $\pT$
in the late hadronic stage.
The difference of  $\langle \pT \rangle$ between with and without  hadronic rescatterings
are relatively smaller for $\pi$, $\phi$, $\Xi$, and $\Omega$ than for the others.
In order to confirm a pure flow effect we also calculate $\langle \pT \rangle$ with option (III)
in which resonance decay contributions are also not included.
As is well-known in analysis using the blast wave model \cite{Schnedermann:1993ws},
 $\langle \pT \rangle$ is a linear function of the hadron mass 
in this case \cite{Heiselberg:1998es}.
We also confirm that it is the case even in our fully dynamical calculations.
$\pT$-averaged $v_{2}$ 
for three options (I), (II), and (III)
are compared with each other in Fig.~\ref{fig:pt&v2_versus_mass} (b). 
By comparing the results (I) and (II) one finds that due to hadronic
  rescatterings $v_{2}$ for hadrons other than pions increases little -- by at
  most $\sim$6\% -- but $v_{2}$ for pions increases by $\sim$25\, 
 even though
  elliptic flow parameters are commonly believed to be sensitive to the
  early stage of the reaction.
 On the other hand, the effect of hadronic
  rescattering on pion mean $p_{T}$ is almost absent \cite{Hirano:2005wx}.
It is interesting to
  note that the effect of resonance decays on pion $v_{2}$ is also significant,
  see Fig.~\ref{fig:pt&v2_versus_mass} (b).
How resonance decays reduce pion $v_{2}$ was discussed in
  Ref.~\cite{Hirano:2000eu}. As we expected, neither $\langle p_{T} \rangle$ 
nor $v_{2}$ for multistrange
  hadrons is affected by hadronic rescatterings, but at least one of
  these observables is affected for all other particles.

\begin{figure*}[tbp]
\begin{center}
\includegraphics[clip,angle=-90,width=0.48\textwidth]{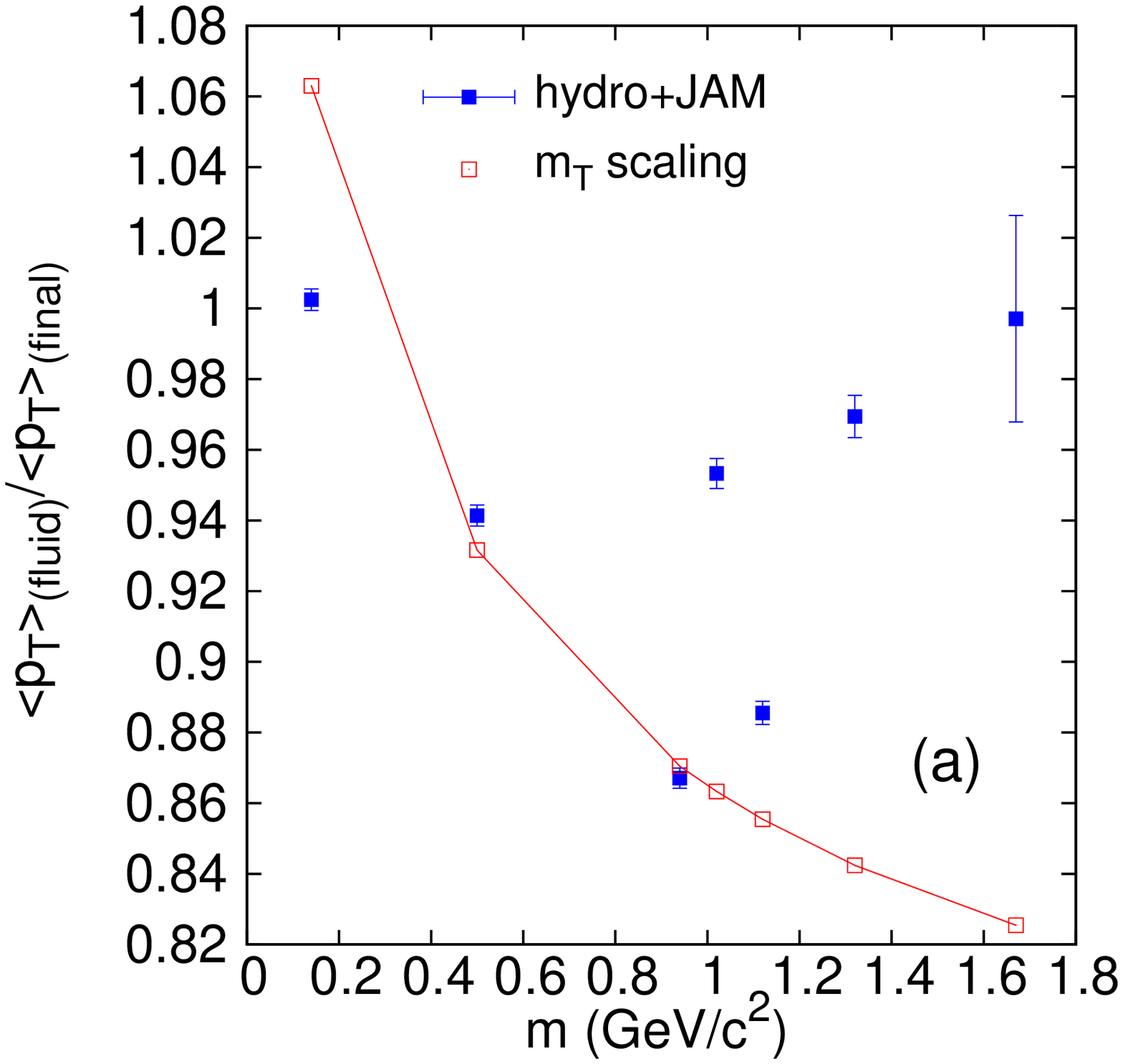}
\includegraphics[clip,angle=-90,width=0.48\textwidth]{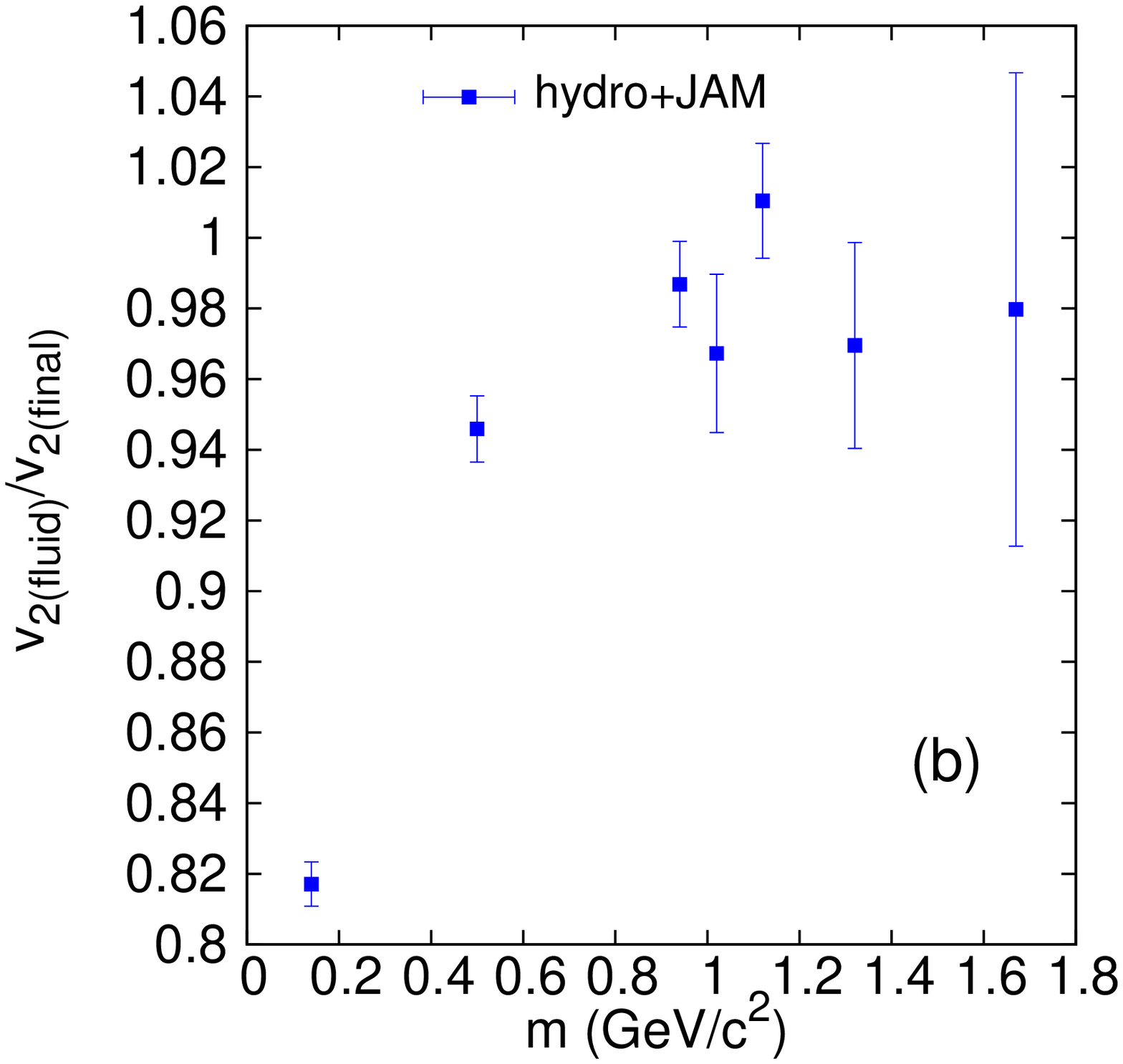}
\end{center}
\caption{(Color Online) Ratio of observables at the fluid stage to the
  final ones in minimum bias $\sNN=200\GeV$ Au+Au collisions in
  $\left| y \right| < 1.0$ obtained by the integrated dynamical
  approach (filled square) for (a) mean transverse momentum
  $\langle \pT \rangle$ and (b) $\pT$-averaged $v_{2}$. In comparison,
  ratio of $\langle \pT \rangle$ obtained from a transverse mass scaling ansatz
  (open square) is also shown (see the text). Each point corresponds to
  a hadron as shown in the caption of Fig.~\ref{fig:pt&v2_versus_mass}.}
\label{fig:ratio}
\end{figure*}

To see how much hadronic rescatterings affect the final observables,
we show ratios of $\langle \pT \rangle$ and $v_{2}$ at 
particlization but after decays (option (II)) to the ones in the final state
(option (I)) \cite{Hirano:2010jg} 
for each identified hadron 
in Fig.~\ref{fig:ratio}.
For comparison,
we also show the ratio  obtained 
from a $m_{T}$ scaling ansatz (open square) \cite{Heiselberg:1998es}
 in the case of mean $\pT$.
In this ansatz, one can parametrize $\pT$ distribution as
\begin{eqnarray}
\frac{dN}{\pT d\pT} &\propto& \exp \left( -{\frac{m_{T}}{T_{\rm eff}}} \right), \\
T_{\rm eff}&=&T_{f} +\frac{1}{2} mv_{f}^{2},\\
m_{T} & = & \sqrt{\pT^2+m^2}, 
\end{eqnarray}
where $T_{f}$, $m$, and $v_{f}$ are
freeze-out temperature, particle mass, and flow velocity, respectively.
Parameters were chosen to 
reproduce the slope of the proton $\pT$ spectrum in option (II) 
($T_{f} =0.155 \GeV$ and $v_{f}=0.56$) and in option (I)
($T_{f} =0.130 \GeV$ and $v_{f}=0.71$).
If all the particles interact strongly with each other 
and move at a common collective velocity
during the hadronic stage,
mass ordering would appear.
In fact, the result from the $m_{T}$ scaling
depends monotonically on particle mass, and
pions, 
kaons, and protons follow this monotonic tendency
obtained from $m_{T}$ scaling ansatz.
However, multistrange hadrons (including ``hidden'' strangeness)
($\phi$, $\Xi$, and $\Omega$) in the integrated dynamical approach
deviate from the pattern.
This means that multistrange hadrons freeze out earlier 
than other hadrons
and do not participate in the radial flow during the transport stage.
Multistrange hadrons couple weakly 
with the system consisting mostly of pions
since they have small scattering cross sections
and hardly rescatter with pions in the late hadronic stage.

Mean transverse momentum $\langle \pT \rangle$ for dominant constituents of the medium, 
namely pions, decreases 
during the hadronic rescattering stage 
since $pdV$ work done in the longitudinal direction 
reduces the transverse energy per unit 
rapidity~\cite{Gyulassy:1983ub,Ruuskanen:1984wv} 
but the number of pions is fixed.
In this way,
\begin{equation}
\label{eq:meanpTineq}
\frac{\langle \pT \rangle_{(\mathrm{fluid})}}{\langle \pT \rangle_{(\mathrm{final})}} \approx
\frac{(1/N_{\pi})(dE_{T}/dy)_{(\mathrm{fluid})}}{(1/N_{\pi})(dE_{T}/dy)_{(\mathrm{final})}}>1.
\end{equation}
In fact, the two parameter sets
 in the $m_{T}$ scaling ansatz mentioned above
are chosen to obey  this inequality (\ref{eq:meanpTineq}).
As for $v_{2}$ in Fig.~\ref{fig:ratio} (b), 
about 20\% of final $v_{2}$ of pions is generated during the late transport
stage.
In other words, only $\sim$80\% of final $v_{2}$ reflects
elliptic flow generated during 
the fluid-dynamical stage. Obviously, this could depend on centrality
(and perhaps collision energy) and the number can be regarded as an average 
value at the RHIC energy.
Whereas hadronic rescatterings affect little final $v_{2}$ for the
other hadrons. 
By combining these two results in Fig.~\ref{fig:ratio},
we conclude that  
multistrange hadrons are less affected by hadronic rescatterings
and, therefore, can be used to probe the hadronization stage
in high-energy nuclear collisions.

Coming back to violation of mass ordering in differential $v_{2}$ shown in 
Fig.~\ref{fig:v2_noscat_scat} and also in Fig.~5 in Ref.~\cite{Hirano:2008prc},
 one can interpret this intriguing phenomenon as follows.
For pions $\pT$-averaged $v_{2}$ increases but $\langle \pT \rangle$ changes little in 
the hadronic stage, 
so slope of ${v_{2}^{\pi}(\pT)}$ gets steeper as seen from 
Eq.~(\ref{eq:v2slope}).
On the other hand, for protons, $\pT$-averaged $v_{2}$ does not change so much, 
but $\langle \pT \rangle$ increases in the hadronic stage,
so ${v_{2}^{p}(\pT)}$ decreases.
However, for $\phi$ mesons, both $\pT$-averaged $v_{2}$ and $\langle \pT \rangle$ 
after 
the hydrodynamic stage are almost the same 
in the final state and, consequently, $v_{2}^{\phi}(\pT)$ is
unchanged.

\begin{figure*}[tbp]
\begin{center}
\includegraphics[clip,angle=-90,width=0.48\textwidth]{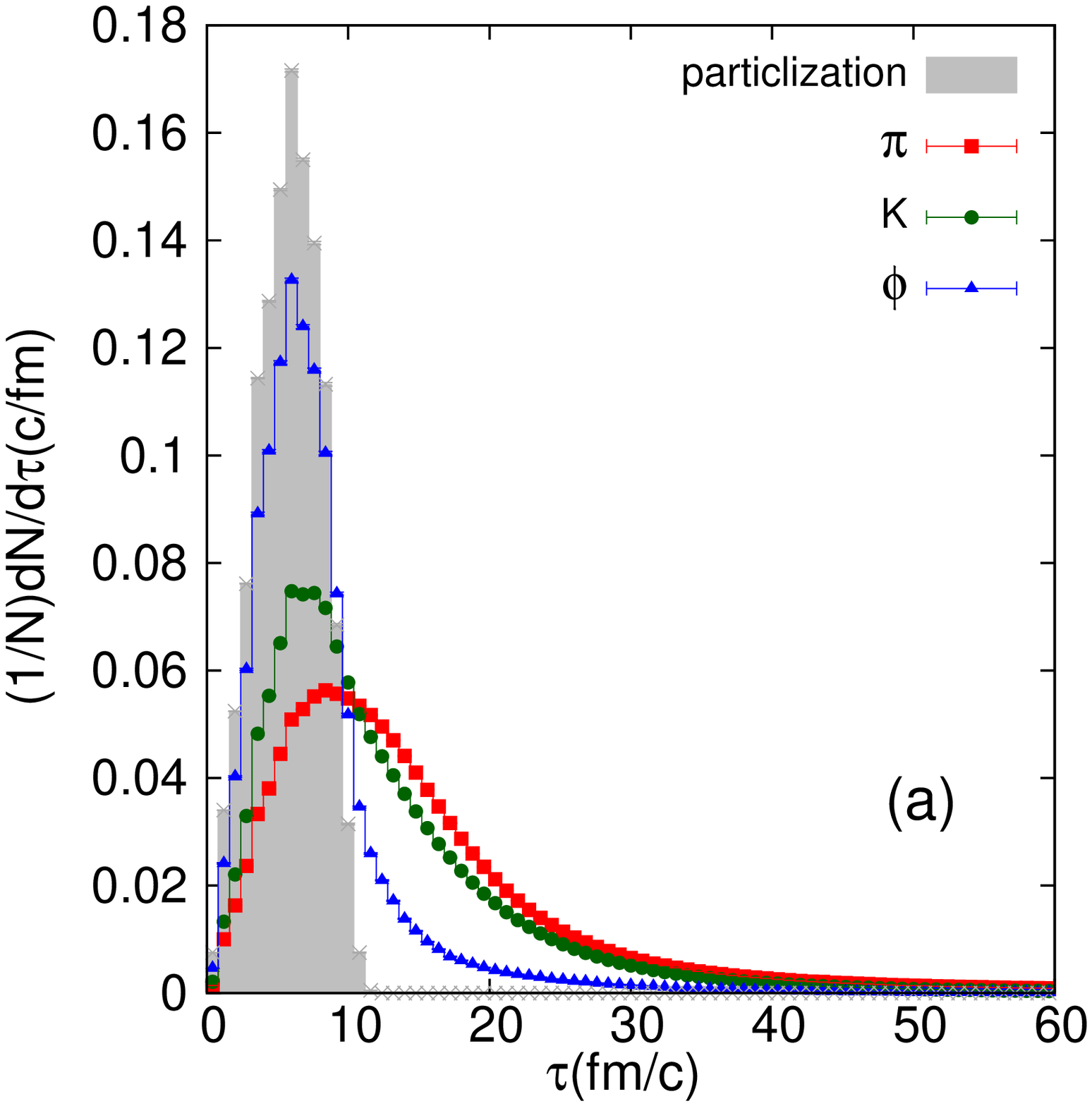}
\includegraphics[clip,angle=-90,width=0.48\textwidth]{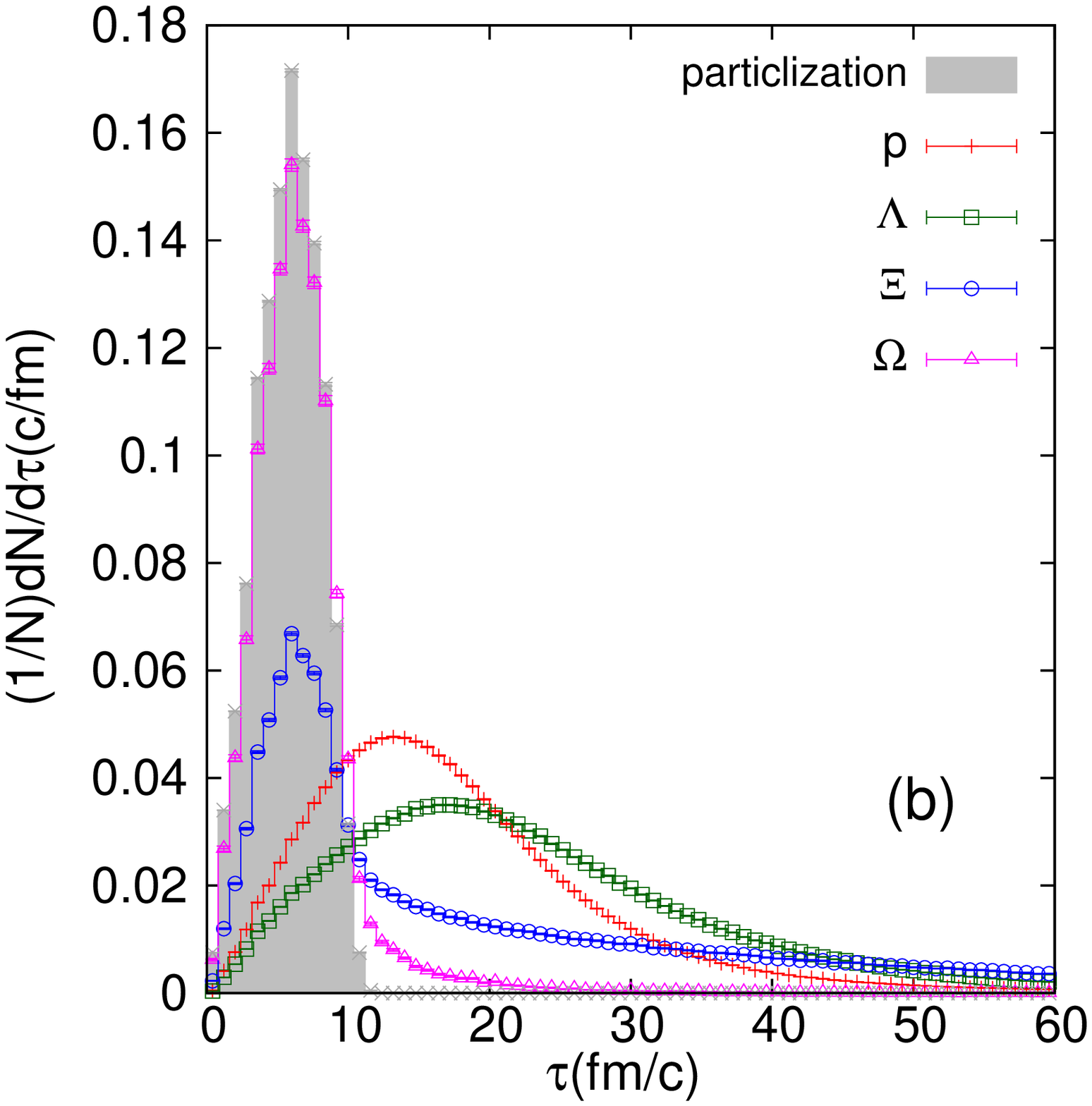}
\end{center}
 \caption{(Color Online) Normalized 
 freeze-out time ($\tau$) distributions
 for (a) mesons: $\pi$, $K$, and $\phi$,
 and (b) baryons: $p$, $\Lambda$, $\Xi$, and $\Omega$
in $\left| y \right| < 1.0$ in minimum bias
 $\sNN=200\GeV$ Au+Au collisions.
The shaded areas correspond to 
normalized particlization time 
distributions of charged hadrons.
}
\label{fig:dNdtau}
\end{figure*} 

Finally we show the normalized 
freeze-out time (defined as the longitudinal proper time $\tau$ of the 
last interaction in hadronic cascade simulations) distributions
around mid-rapidity ($\left| y \right| < 1.0$) 
in minimum bias Au+Au collisions at $\sNN=200\GeV$.
In Fig.~\ref{fig:dNdtau} the results are shown in two panels, 
(a) for mesons and (b) for baryons, for clarity.
The shaded areas represent the particlization time distribution for
  charged hadrons. Since the particlization time does not depend on
  scattering cross sections, it is almost equal for all hadrons within
  error bars. A prominent peak with small width is seen below $\tau = 10$
  fm/$c$, corresponding to the time of particlization at different parts of
  the system. The freeze-out time distributions of $\phi$ and $\Omega$ do not
  differ from the particlization time distribution significantly, which
  reflects their small scattering cross sections. On the other hand, other
  hadrons have broader freeze-out time distributions. 
For $\Xi$, a peak also exists at an early time like for $\phi$ and $\Omega$
but its height is lower and the peak is accompanied by a broad tail.
This is because $\Xi$ has a contribution from long-living resonance $\Xi$(1530) 
(decay width $\Gamma =9.9$ MeV $\sim$ 1/(20  fm/$c$) \cite{PDG:2014CPC}) 
whose decay forms a long tail in the freeze-out time distribution.
So the primordial $\Xi$ freezes out as early as $\phi$ and $\Omega$ do.
These results support our findings above 
that multistrange hadrons freeze 
out just after the particlization due to less rescatterings in the 
late kinetic stage.

\section{Summary}
\label{sec:summary}
We have studied the effects of the hadronic rescatterings 
especially on multistrange hadrons 
and claimed they can be used to probe the hadronization stage
in high-energy nuclear collisions.
We have employed an integrated dynamical model in which the ideal hydrodynamic model is
 combined with the hadronic cascade model, JAM, and confirmed that our model fairly 
reproduces the experimental data of $\pT$ spectra and differential $v_{2}$ not only for 
pions, kaons, and protons but also for multistrange hadrons.
Within this approach, we have simulated the hadronic stage with or
without the rescatterings and compared the calculated 
mean transverse momentum and $\pT$-averaged $v_{2}$ in these two cases
to investigate the hadronic rescattering effects.
We have found that the multistrange hadrons are less affected by the 
rescatterings than non-strange hadrons.
Because of the small scattering cross sections,
the multistrange hadrons do not fully participate in the radial flow during
the hadronic stage and freeze out earlier than non-strange hadrons.
With these results we have shown 
how to interpret behaviors
 of mass ordering and 
its violation in $v_{2}(\pT)$.
Changes of slope of $v_{2}(\pT)$ during the hadronic rescattering stage
result from interplay between
changes of mean $\pT$ and $\pT$-averaged $v_2$.
We have also showed the freeze-out time distributions for identified hadrons and confirmed 
that the multistrange hadrons freeze out soon after they are ``particlized''.

In future, we also plan to investigate the hadronic rescattering effects
 on observables for multistrange hadrons
at the LHC energy. Although
elliptic flow parameters
 are expected to be less affected
by the hadronic rescatterings at the LHC, 
 it would be important to quantify how much they do change, and thus how 
much the observed $v_2$ at LHC reflects the QGP evolution.

\begin{acknowledgements}
We acknowledge a fruitful discussion with N.~Xu.
This work was supported by JSPS KAKENHI Grant Numbers 
12J08554(K.M.) and 25400269 (T.H.), and by BMBF under contract no. 06FY9092(P.H.).

\end{acknowledgements}

\end{document}